\documentclass[prb,twocolumn,showpacs]{revtex4}

\usepackage{amssymb,amsmath,bm,epsfig,graphics,subfigure,multirow}
\newcommand{\br}[1]{\mathbf{#1}}
\newcommand{\bg}[1]{\boldsymbol{#1}}
\newcommand{\ten}[1]{\overleftrightarrow{#1}}
\newcommand{\ax}{x_{\tiny{\mbox{$\|$}}}}
\newcommand{\cu}{x_{\tiny{\mbox{$\bot$}}}}
\newcommand{\qa}{q_{\tiny{\mbox{$\|$}}}}
\newcommand{\qc}{q_{\tiny{\mbox{$\bot$}}}}
\newcommand{\ga}{g_{\tiny{\mbox{$\|$}}}}
\newcommand{\gc}{g_{\tiny{\mbox{$\bot$}}}}
\newcommand{\Gcc}{G_{\tiny{\mbox{$\bot$}}\tiny{\mbox{$\bot$}}}}
\newcommand{\Gaa}{G_{\tiny{\mbox{$\|$}}\tiny{\mbox{$\|$}}}}
\newcommand{\Gac}{G_{\tiny{\mbox{$\|$}}\tiny{\mbox{$\bot$}}}}
\newcommand{\Gca}{G_{\tiny{\mbox{$\bot$}}\tiny{\mbox{$\|$}}}}

\begin{document}

\title{Casimir Interactions Between Scatterers in Metallic Carbon Nanotubes}
\author{Dina Zhabinskaya}
    \email{dinaz@physics.upenn.edu}
 \author{E.J. Mele}   
    \affiliation{Department of Physics and Astronomy \\ University of Pennsylvania, Philadelphia PA 19104}
\date{\today}
\begin{abstract}
We study interactions between localized scatterers on metallic carbon nanotubes by a mapping onto a one-dimensional Casimir problem. Backscattering of electrons between localized scattering potentials mediates long range forces between them. We model spatially localized scatterers by local and non-local potentials and treat simultaneously  the effects of intravalley and intervalley backscattering.  We find that the long range forces between scatterers exhibit the universal power law decay of the Casimir force in one dimension, with prefactors that control the sign and strength of the interaction. These prefactors are nonuniversal and depend on the symmetry and degree of localization of  the scattering potentials. We find that local potentials inevitably lead to a coupled valley scattering problem, though by  contrast non-local potentials lead to two decoupled single-valley problems in a physically realized regime.  The Casimir effect due to two-valley scattering potentials is characterized by the appearance of spatially periodic modulations of the force.
\end{abstract}
\pacs{03.70.+k,73.63.,11.80.-m,61.72.S-}
\maketitle 
\section{Introduction} 

A single-walled carbon nanotube is a two dimensional graphene sheet rolled into a cylinder. The diameter of the nanotube is on the order of a few nanometers, and its length can vary from hundreds of nanometers to centimeters. Due to the small tube radius, electrons are confined in the azimuthal direction, and at sufficiently low energy the quantum confinement leads to an effectively  one-dimensional electronic system. These nanotubes can be either metallic or semiconducting, and the low-energy electronic band structure can be studied using a long-wavelength expansion of the Hamiltonian around each of the degenerate Fermi points, labeled by K and K' points. This long-wavelength theory is given by a pair of one dimensional Dirac Hamiltonians. 

When a nanotube is chemically functionalized or contains defects on the tube wall, localized scattering centers interrupt the free motion of its low-energy charge carriers. Generally a localized defect can backscatter a propagating low-energy electron, either by large momentum scattering between the K and K' valleys or by small momentum backscattering from forward to backward moving excitations within a single valley. Superposition of right and left moving excitations produces various standing wave patterns in the electron density near such a defect. 

In this paper we consider forces on the scatterers produced by their interaction. It is easy to see that for an isolated scatterer, the backscattering-induced forces on the left and right hand side of the defect must exactly cancel, so there is no net force. However, for pairs of defects and generally for a distribution of defects at finite density, the forces on the left and right hand sides of the scatterer do not balance and mediate a net force on each scatterer. In previous work we explored this effect within a single valley model for the nanotube and found that the scattering induced forces could be mapped to a Casimir-type problem, where the propagating electron waves provide the role of the background quantum field. Importantly, the spinor character of these background fermions admits the possibility of attractive, repulsive or compensated null forces on the scatterers depending on the internal symmetry of their scattering potentials [\onlinecite{first.paper}]. 

In this work we generalize these earlier results to study the combined effects of intravalley and intervalley backscattering. This extension proves to be crucial for a meaningful application to the nanotube problem. Potentials that produce only intravalley scattering need to vary slowly on the scale of a lattice spacing. Yet, any local potential with this property degenerates to a one-dimensional scalar potential that cannot backscatter a massless Dirac particle. Thus, for a local potential our effect ultimately requires a significant degree of spatial localization, and in this regime intervalley backscattering ultimately arises. Indeed, we find below that for local potentials there is no regime in which the force problem can be regarded as confined to a single valley, necessitating a coupled valley formulation of the scattering problem.  

By contrast, non-local scattering potentials do allow the possibility of only intravalley backscattering in a controlled physically realizable limit. This situation is realized most naturally for electrons coupled to slowly varying lattice strains on a nanotube.  In this paper we present a generalization of the formalism described in Ref. [\onlinecite{first.paper}] suitable for application to the coupled two valley problem, and explore the forces that occur as a function of range and internal symmetry of the scattering potentials.  We provide formulae that describe the electron mediated forces in these various geometries. Table I provides a compact summary of our results.

The magnitude and sign of the interaction is dictated by the internal structure of the scatterers.  Local potentials can describe atomically sharp impurities localized on a sublattice site.
We find a repulsive force between local impurities residing on equivalent sublattice sites and an attractive force between scatterers on distinct sites.
Related results were recently shown for interactions between impurities in two-dimensional graphene [\onlinecite{Levitov}].
We also explore interactions between impurities where only intervalley scattering is present.  Interactions between defects due to large momentum backscattering were previously discussed in one-dimensional Fermi liquids [\onlinecite{Recati2},\onlinecite{liquids}].
For non-local potentials we show that scattering persists for ranges that are larger than the lattice constant leading to the single-valley scattering problem. The results we obtain for Casimir forces between non-local scatterers agree with our previous work. We recover the universal distance dependent power law decay for the Casimir force in one-dimension.
However, for local potentials, unlike for the single-valley problem, we also observe periodic spatial modulations in the force due to intervalley scattering.

The paper is organized in the following manner.  In Sec.~\ref{sec:Ho} we define the geometry and derive the low-energy electronic structure of single-walled carbon nanotubes. In Sec.~\ref{sec:H1} we present scattering potentials which can describe impurities in nanotubes.  The distinction between relevant length scales is discussed in Sec.~\ref{sec:range}. In Sec.~\ref{sec:local} and Sec.~\ref{sec:non-local} we discuss local and non-local potentials, respectively. In Sec.~\ref{sec:force} we outline the basic mechanism used to calculate Casimir forces.  In Sec.~\ref{sec:1valley} we review our previous work of the one-valley problem, and show how the method is generalized to the two-valley problem in Sec.~\ref{sec:2valley}.  
  Our main results are presented in Sec.~\ref{sec:results}. Casimir forces between local and non-local potentials are shown in Sec.~\ref{sec:Flocal} and  Sec.~\ref{sec:Flocal}, respectively.  In Sec.~\ref{sec:dis} we discuss the relation of our findings to physical adsorbates on nanotubes.
  The paper is concluded in Sec.~\ref{sec:conclusion}.

\section{Single-Walled Carbon Nanotubes} \label{sec:Ho}
\subsection{Nanotube Geometry}
\begin{figure*}[t]
\includegraphics*[width=5.8in]{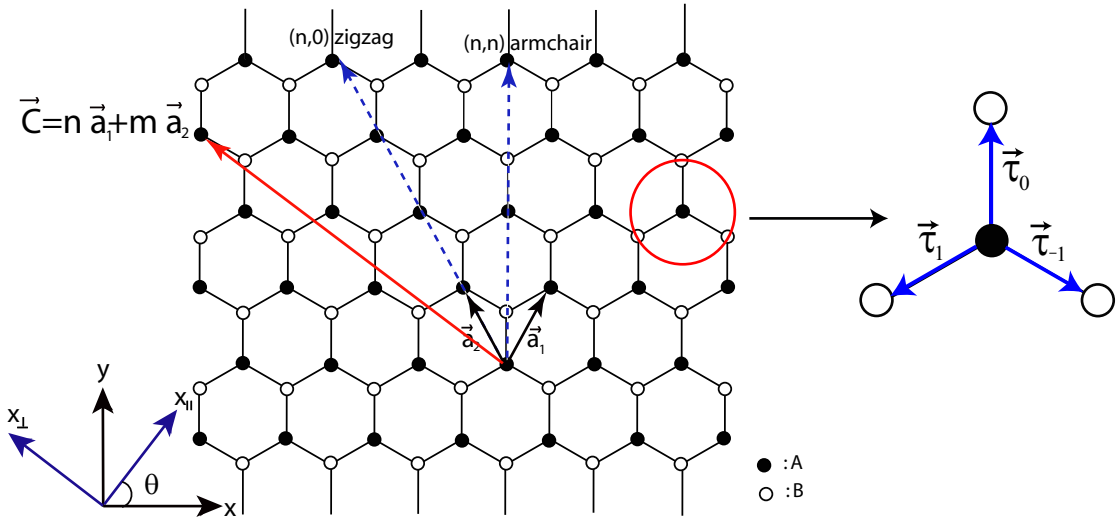}
\caption{Two-dimensional honeycomb lattice with $A$ and $B$ sublattice sites identified. 
The primitive unit vectors are $\br{a}_1$ and $\br{a}_2$, and $\bg{\tau}'s$ define a triad of nearest neighbor bond vectors.  A nanotube is characterized by a vector $\br{C}=n\br{a}_1+m\br{a}_2$ that point along the tube circumference.  The chiral angle $\theta$ is the angle between the lattice coordinate $x$ and the tube axis $\ax$.  The circumference vectors of high-symmetry achiral armchair (n,n) and zigzag (n,0) nanotubes are shown.}
\label{fig:lattice}
\end{figure*}

In this section we describe the geometric structure of a carbon nanotube and introduce the notation used in this paper. 
Two-dimensional graphene is a honeycomb lattice with two inequivalent sublattice sites, labeled $A$ and $B$, as illustrated in Fig.~\ref{fig:lattice}.  There is one carbon atom residing on each lattice site.  The primitive lattice vectors are $\br{a}_1=a(1,\sqrt{3})/2$ and $\br{a}_1=a(-1,\sqrt{3})/2$, where $a/\sqrt{3}\sim 1.4$~\AA~is the nearest neighbor bond length.  The vectors $\bg{\tau}$'s define a triad of nearest neighbor bond vectors as shown in Fig.~\ref{fig:lattice}.

A carbon nanotube is formed by wrapping the graphene sheet into a cylinder, such that two equivalent lattice sites are identified.  The circumferential vector $\br{C}=n\br{a}_1+m\br{a}_2$, where $n,m\in\mathbb{Z}$, characterizes the nanotube.  The $xy$-plane defines 
the lattice coordinate system, where bonds run parallel to the $y$-axis.  The tube coordinate system is defined by $\ax$ along the tube axis and $\cu$ around the circumference. The two coordinate systems 
are related by the tube's chiral angle defined as the angle between $x$ and $\ax$ as shown in Fig.~\ref{fig:lattice}.  The coordinate transformation is given by
\begin{equation}
\begin{pmatrix}
\hat{x}_{\tiny{\mbox{$\|$}}} \\ \hat{x}_{\tiny{\mbox{$\bot$}}} \\ 
\end{pmatrix}
=
\begin{pmatrix}
\cos\theta & \sin\theta \\ -\sin\theta & \cos\theta \\
\end{pmatrix}
\begin{pmatrix}
\hat{x} \\ \hat{y} \\ 
\end{pmatrix}
.
\label{eq:euler}
\end{equation}
The circumference vectors of high-symmetry achiral nanotubes that have a plane of mirror symmetry are shown in Fig.~\ref{fig:lattice}. In armchair ($\theta=0$) and zigzag ($\theta=\pi/6$) carbon nanotubes bonds run parallel to the tube's circumference and axis, respectively. 

Fixing the origin on an $A$ site, the lattice translation vector $\br{R}_A=n_1\br{a}_1+n_2\br{a}_2$, where $n_1,n_2\in\mathbb{Z}$, locates an $A$ sublattice site, and the vector $\br{R}_B=\br{R}_A+\bg{\tau}_o$ locates a $B$ site, where $\bg{\tau}_o$ is a vector connecting the two sublattice sites.
The lattice vectors in the nanotube coordinate system are given by
\begin{align}
\br{R}_i&=\frac{a}{2}\Big[\cos\theta\Big(n_1-n_2\Big)+\sqrt{3}\sin\theta\Big(n_1+n_2+\frac{2b}{3}\Big)\Big]\hat{x}_{\tiny{\mbox{$\|$}}}\nonumber \\
&+\frac{a}{2}\Big[-\sin\theta\Big(n_1-n_2\Big)+\sqrt{3}\cos\theta\Big(n_1+n_2+\frac{2b}{3}\Big)\Big]\hat{x}_{\tiny{\mbox{$\bot$}}},
\label{eq:lattice.vec}
\end{align}
where $b=0$ for $i=A$ and $b=1$ for $i=B$.
The nearest neighbor bond vectors $\bg{\tau}_{j}$'s shown in Fig.~\ref{fig:lattice} in the tube coordinate system are given by
\begin{equation}
\bg{\tau}_{j}=\frac{a}{\sqrt{3}}\Big(\sin\theta_{j}\hat{x}_{\tiny{\mbox{$\|$}}}+\cos\theta_{j}\hat{x}_{\tiny{\mbox{$\bot$}}}\Big),
\end{equation}
where $\theta_{j}=\theta-2\pi j/3$, and $j=\{0,\pm 1\}$.

\begin{figure}[h]
\includegraphics*[width=2.8in]{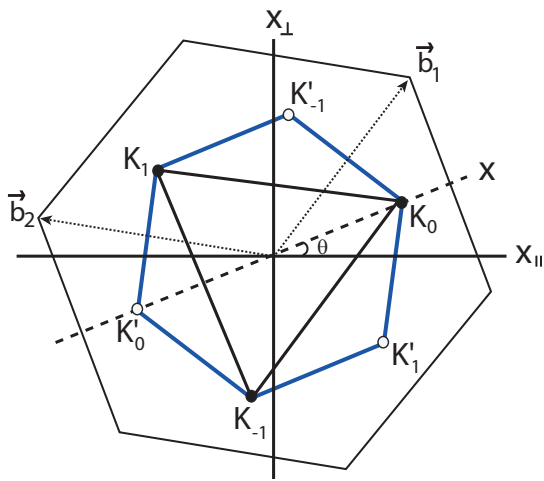}
\caption{The first Brillouin zone of the honeycomb lattice depicted relative to the tube coordinate system, where $\ax$ points along the tube axis.  The six corners of the Brillouin zone are shown.  The three equivalent $K$ (black circles) and $K^{\prime}$ (white circles) points are related by reciprocal lattice vectors  $\br{G}=m_1\br{b}_1+m_2\br{b}_2$.  The chiral angle $\theta$ is defined as the angle between the tube axis and $\br{K}_0$ in the lattice coordinate system.} 
\label{fig:star}
\end{figure}

The first Brillouin zone of the honeycomb lattice is shown in Fig.~\ref{fig:star}.  In graphene the conduction and valence bands touch at the six corner points of the Brillouin zone.  Therefore, for undoped graphene the Fermi surface lies at the $K$ and $K^{\prime}$ points. The three equivalent Fermi points identified by white and black circles in Fig.~\ref{fig:star} are related by reciprocal lattice vectors $\br{G}=m_1\br{b}_1+m_2\br{b}_2$.  However, $K$ and $K^{\prime}$ points are inequivalent since they cannot be connected through a reciprocal lattice vector.  In the nanotube coordinate system the six corners of the Brillouin zone are given by
\begin{equation}
\alpha\br{K}_p=\alpha\frac{4\pi}{3a}\Big(\cos\theta_p\hat{x}_{\tiny{\mbox{$\|$}}}-\sin\theta_p\hat{x}_{\tiny{\mbox{$\bot$}}}\Big),
\label{eq:Kpoint}
\end{equation}
where $\alpha=+1(-1)$ for $K(K^{\prime})$-points, and $p=\{0,\pm 1\}$.  As shown in Fig.~\ref{fig:star}, the corner point $\br{K}_o$ is a reference defining the chiral angle $\theta$ between the lattice $x$-axis and the tube axis.  
  
\subsection{Low-Energy Theory}

The energy band structure of graphene can be obtained using a tight-binding model for $\pi$ electrons.  Considering nearest neighbor hopping between sites on a two-dimensional honeycomb lattice, the tight-binding Hamiltonian for graphene is given by
\begin{equation}
\mathcal{H}_o=-t\sum_{\br{R}_A,j}a^{\dagger}(\br{R}_A)b(\br{R}_A+\bg{\tau}_j)+h.c.,
\label{eq:tb}
\end{equation}
where $t$ is the nearest neighbor hopping energy, and $a^{\dagger}(b^{\dagger})$ creates an electron on the A(B) sublattice.  The low-energy electronic properties are found by expanding the tight-binding Hamiltonian around the two distinct  $K(K^{\prime})$-points to linear order in momentum $\br{k}$.  Since the Fermi points are inversely proportional to the lattice constant $|\br{K}|\propto 1/a$, the long-wavelength theory is valid for $|\br{k}|a\ll 1$.  

The energy spectrum of a carbon nanotube is obtained from the graphene Hamiltonian by rotating to the tube coordinate system and quantizing the crystal momentum along the transverse direction.
Single-walled carbon nanotubes are either metallic or semiconducting depending on whether the discrete lines of crystal momentum pass through the Fermi points $K$ and $K^{\prime}$.  It turns out that $1/3$ of all nanotubes are metallic, since mod$(n-m,3)=0$ is a necessary condition for the six corners of the Brillouin zone to be allowed wave vectors.  

In our notation, the 
$2\times 2$ identity and Pauli matrices $\{I_{\sigma},\sigma_{i}\}$ span $A(B)$-sublattice pseudospin space, and $\{I_{\tau},\tau_i\}$ span the $K(K^{\prime})$-point valley isospin space, where $i=\{x,y,z\}$. For simplicity, we introduce an operator which defines a rotation by an angle $\eta$ around $\br{\hat{n}}$
in either $\tau$ or $\sigma$ space.
For example, in $\sigma$ space this operator is given by
\begin{equation}
\mathcal{O}^{\prime}(\br{\hat{n}}_{\sigma},\eta)\equiv e^{i\br{\hat{n}}\cdot\sigma\eta/2}\mathcal{O}e^{-i\br{\hat{n}}\cdot\sigma\eta/2}.
\end{equation}
It is convenient to define a projection operator $P^{\pm}_{\sigma}=(I_{\sigma}\pm\sigma_z)/2$ which projects on a sublattice site.  Likewise, $P^{\pm}_{\tau}=(I_{\tau}\pm\tau_z)/2$ is a projection operator in the valley space.

In this paper, we only consider the lowest energy band of metallic tubes (gapless systems) as will be explained in Sec.~\ref{sec:H1}. Expanding $\mathcal{H}_o$ around the Brillouin zone corners and rotating to the tube coordinate system using Eq.~(\ref{eq:euler}), the long-wavelength Hamiltonian for the lowest energy band of a metallic nanotube becomes
\begin{align}
\Big(-i\hbar v_F\Big[&P^{+}_{\tau}\otimes\sigma^{\prime}_x(\hat{z}_{\sigma},-\theta_p)
-P^{-}_{\tau}\otimes\sigma^{\prime}_x(\hat{z}_{\sigma},\theta_p^{\prime}) \Big]\partial_{\ax}\nonumber \\
&-E\Big)f_k(\ax) =0,
\label{eq:Ho}
\end{align}  
where $\hbar v_F=\sqrt{3}at/2\sim 0.54$~eV$\cdot$nm.  The basis states are four-component spinors defining relative amplitudes at the $A$ and $B$ sites and the $K$ and $K^{\prime}$ Fermi points in the following order
$(AK_p,BK_p,AK^{\prime}_{p^{\prime}},BK^{\prime}_{p^{\prime}})$, where $p$ and $p^{\prime}$ correspond to one of the three equivalent $K$ and $K^{\prime}$ points, respectively, depicted in Fig.~\ref{fig:star}. The eigenstates of $\mathcal{H}_o$, $f^{\alpha p}_{\pm k}(\ax)=\phi^{\alpha p}_{\pm k}e^{\pm ik\ax}/\sqrt{2\pi}$ are right and left moving plane waves multiplied by a spinor, where $k$ is the momentum along the tube axis.  When the chemical potential is fixed at $\mu =0$ the filled Dirac sea has $E=-|k|$, and the right and left moving spinors are given by
\begin{equation}
\phi^{p}_{\pm k}=\frac{1}{\sqrt{2}}
\begin{pmatrix}
1 \\ \mp e^{i\theta_p} \\ 0 \\ 0 \\
\end{pmatrix}
,~~~\phi^{-p^{\prime}}_{\pm k}=\frac{1}{\sqrt{2}}
\begin{pmatrix}
0\\ 0 \\ 1 \\ \pm e^{-i\theta_p^{\prime}} \\
\end{pmatrix}
.
\label{eq:spinors}
\end{equation}

\subsection{Basis States} \label{sec:basis}
The eigenstates of the long-wavelength Hamiltonian in Eq.~(\ref{eq:Ho}) are isotropic and do not depend on the crystal orientation of a nanotube.  To include lattice anisotropic potentials in the theory, we reconstruct the Bloch functions from the solutions in Eq.~(\ref{eq:Ho}) for the effective mass theory.
In the $\br{k}\cdot\br{p}$ approximation the electron wave function near the Fermi energy is given by a Bloch function at the $K$ point multiplied by an envelope function.  For graphene, the wave function is
\begin{equation}
\Psi(\br{K}+\br{k},\br{r})=\sum_{i=A,B}e^{i\br{k}\cdot\br{r}}\Psi_{i,\br{K}}(\br{r})\phi_{i,\br{k}},
\end{equation}
where $\Psi_{i,\br{K}}(\br{r})$'s are exact Bloch functions at the $K$ point, and $e^{i\br{k}\cdot\br{r}}\phi_{i,\br{k}}$'s are slowly varying envelope functions [\onlinecite{Mele.mass}]. 
Bloch states are plane waves multiplying a cell periodic function.   Potentials which resolve the lattice structure couple to the lattice periodic component of the Bloch states.
Taking the  Fourier transform of the periodic part of the Bloch function, the sublattice basis functions at any of the six corner points $\alpha\br{K}_p$'s are given by
\begin{align}
\Psi^{\alpha p}_{i}(\br{r})&=e^{i\alpha\br{K}_p\cdot\br{r}}u_{i}(\br{r}) \nonumber \\
&=e^{i\alpha\br{K}_p\cdot\br{r}}\sum_nF(|\alpha\br{K}_p+\br{G}_n|)e^{i\br{G}_n\cdot(\br{r}-\bg{\tau}_i)},
\label{eq:basis}
\end{align}
where $F(q)$ is the Fourier transform of a localized orbital function, $\br{G}$'s are reciprocal lattice vectors, and $\alpha$ defined in Eq.~(\ref{eq:Kpoint}) labels the $K$ and $K^{\prime}$ points.  The subscript $i$ labels a sublattice site, such that $\bg{\tau}_A=0$ and $\bg{\tau}_B=\bg{\tau}_o$.   The functions in Eq.~(\ref{eq:basis}) are rapidly oscillating and describe modulations on the scale of the atomic spacing.  Since $F(q)$ decreases rapidly with momentum, in the lowest ``star'' approximation [\onlinecite{first.star}] we keep terms in the sum of $\br{G}_n$'s which connect the three $\br{K}(\br{K}^{\prime})$ Brillouin zone corners, such that $|\br{K}+\br{G}|=|\br{K}|$.  This approximation is appropriate for the range of the scattering potentials we study in this paper.  The normalized basis functions at the $A$ and $B$ sites in the lowest ``star'' representation are given by
\begin{align}
& \Psi^{\alpha p}_A(\br{r})=\frac{1}{\sqrt{3}}\sum_{m=0,\pm 1}e^{i\alpha\br{K}_m\cdot\br{r}} \nonumber \\
& \Psi^{\alpha p}_B(\br{r})=\frac{1}{\sqrt{3}}z^{\alpha p}\sum_{m=0,\pm 1}e^{i\alpha\br{K}_m\cdot\br{r}}z^{-\alpha m},
\label{eq:states.star}
\end{align}
where $z=\exp(i2\pi/3)$.  Evaluating the matrix element of a tight-binding potential in the lowest ``star'' basis given in Eq.~(\ref{eq:states.star}) and expanding to linear order in $\br{k}$, one obtains the low-energy Hamiltonian given in Eq.~(\ref{eq:Ho}) for the lowest band of a metallic nanotube.

\section{Scattering Potential} \label{sec:H1}
In this paper, we study Casimir interactions between two scatterers mediated by the conduction electrons of a carbon nanotube. 
In this section we describe the structure of the scattering potentials used to study this problem.  We explore the dependence of the Casimir interaction on the symmetry, range, strength, and orientation of the two potentials.  We discuss two types of potentials, local and non-local, which result in different scattering processes.   

\subsection{Potential Range} \label{sec:range}
    
In our previous work we studied the one-valley scattering problem valid for potentials whose range is larger than the lattice constant, where intervalley scattering does not play a role.  The $2\times 2$ matrix structure of such a potential is described by its pseudospin polarization [\onlinecite{first.paper}]. When the range of the potential is on the order of interatomic spacing, the two valleys are no longer decoupled [\onlinecite{Ando}]. In this paper, we build upon our previous work to incorporate the effects of sharper potentials resulting in a two-valley scattering problem.  When the two valleys are coupled, the potential is described by a $4\times 4$ matrix and is characterized by both pseudospin and valley polarizations.

\begin{figure}[h]
\includegraphics*[width=3in]{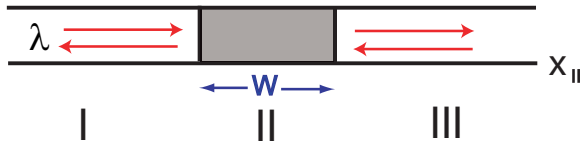}
\caption{An illustration of a scattering process. I and III define regions of free propagation along the tube axis.  The shaded scattering region has a width $W$. A scattering potential can be represented by a delta-function when $W$ is much smaller than $\lambda$, the wavelength of the envelope function.} 
\label{fig:length}
\end{figure}

In general, the spatial variation $W$ of the scattering potential relevant for Casimir interactions is shorter than the conduction wavelength of the envelope function  $\lambda$, such that $Wk\ll 1$.  Fig.~\ref{fig:length} shows an illustration of a scattering process.  Freely propagating electrons in regions I and III have a wavelength $\lambda\propto 1/k$, and the scattering region II has a width $W$.  A potential can be described by delta-function as long as $W\ll \lambda$.  The important distinction between the one- and two- valley scattering problems described by the spinor structure of the Hamiltonian is relevant for potentials whose range is longer and shorter, respectively, than the interatomic separation. 

We study interactions between scatterers in metallic nanotubes.  Since Casimir interactions mediated by massive fields are exponentially suppressed at long distances [\onlinecite{casimir.book}], in this paper we do not address semiconducting nanotubes or scattering between bands which do not pass through the Fermi energy.  The momentum transfer in the azimuthal direction between various Fermi points is determined by the matrix structure of the scattering potential $\hat{V}$.  The free degree of freedom is the longitudinal momentum, and the scattering process is truly one-dimensional along the tube axis.  
We model a delta-function scatterer by a one-dimensional square-barrier potential of the form 
\begin{equation}
\hat{V}(\ax)=\hat{V}\theta(\ax-x_1)\theta(x_2-\ax),
\label{eq:square}
\end{equation}
where $\hat{V}$ describes the internal structure of the potential, and $W=(x_2-x_1)\ll\lambda$ is the barrier width [\onlinecite{first.paper}].

In the rest of the paper, long-range potentials imply a range $d$ longer than the lattice constant but shorter than the envelope function wavelength $a<d\ll\lambda$.  Short-range potentials refer to atomically sharp scatterers whose range is comparable to or smaller than the lattice constant $d\lesssim a$.   

\subsection{Local Potentials} \label{sec:local}
A local potential can be represented as
\begin{equation}
V(\br{r},\br{r}^{\prime})=V(\br{r})\delta(\br{r}-\br{r^{\prime}}).
\label{eq:local}
\end{equation}
We are interested in the matrix structure of the scattering potential as a function of its range and position on the lattice.  For example, if we consider Gaussian model potential $V(\br{r})=Ve^{-|\br{r}-\br{r_o}|/d^2}$, then on a surface of a cylinder $V(\br{r})$ is given by
\begin{equation}
V(\ax,\cu)=V\exp\Bigg\{-\frac{(\ax-\ax^o)^2}{d^2}-\frac{4R^2}{d^2}\sin^2\Big(\frac{\cu-\cu^o}{2R}\Big) \Bigg\},
\label{eq:Gaussian}
\end{equation}
where $V$ is the potential strength, $\br{r_o}=(\ax^o,\cu^o)$ is the center of the Gaussian on the nanotube surface, $R$ is the radius of the tube, and $d$ controls the range of the potential.

  The matrix elements are calculated in the lowest ``star'' basis defined in Sec.~\ref{sec:basis}.  
For example, the intravalley matrix expectation value $V_{AA}$ of the potential given in Eq.~(\ref{eq:local}) evaluated in the lowest ``star'' basis defined in Eq.~(\ref{eq:states.star}) is given by
\begin{align}
\langle\Psi^{p}_A(\br{r})|V(\br{r})|\Psi^{p}_A(\br{r})\rangle=&\frac{1}{3}\sum_{m,m^{\prime}}\int d^2r e^{-i(\br{K}_m-\br{K}_{m^{\prime}})\cdot\br{r}}V(\br{r})\nonumber \\
=&\frac{1}{3}\sum_{m,m^{\prime}}V(\br{K}_m-\br{K}_{m^{\prime}}),
\end{align} 
where $V(\br{q})$ is the Fourier transform of the potential.
The  Fourier transform of the Gaussian potential in Eq.~(\ref{eq:Gaussian}) is normalized such that $V(\br{q})\to 1$ as $\{\qa,\qc\}\to 0$.  Therefore, $V(\qa,\qc)$ is given by
\begin{equation}
V(\qa,\qc)=V\Bigg[I_{\qc R}\Big(\frac{2R^2}{d^2}\Big)\Big{/}I_o\Big(\frac{2R^2}{d^2}\Big)\Bigg]e^{-\qa^2d^2/4}e^{-i\br{q}\cdot\br{r}_o},
\label{eq:Fourier}
\end{equation}
where $I_n(x)$ is a modified Bessel function of the first kind.  In the large radius limit, the Fourier transform of the Gaussian potential approaches the limit of a potential on a two-dimensional flat sheet and becomes isotropic.  In the  $R\gg a$ limit Eq.~(\ref{eq:Fourier}) is given by
\begin{equation}
V(\br{q})=Ve^{-|\br{q}|^2d^2/4}e^{-i\br{q}\cdot\br{r}_o}.
\label{eq:2d}
\end{equation}
We define the center of the Gaussian by $\br{r}_o=\br{R}_A^o+\nu\bg{\tau}_{\ell}$, where $0\leq\nu\leq 1$, such that the potential is centered on either the $A$ sublattice, the $B$ sublattice, or along any of the three bonds defined by the triad of bond vectors $\bg{\tau}_{\ell}$ pointing away from $\br{r}_A^o$. 
The total impurity Hamiltonian is given by
\begin{equation}
\mathcal{H}_1=\mathcal{H}^a_1+\mathcal{H}^e_1,
\label{eq:Hlocal}
\end{equation}
where $\mathcal{H}^a_1$ and $\mathcal{H}^e_1$ are $4\times 4$ matrices containing $\it{intravalley}$ and $\it{intervalley}$ matrix elements, respectively.

Initially, we focus on the intravalley part of the potential. Evaluating both the diagonal and off-diagonal  matrix elements $V_{ij}$'s of a local potential, the intravalley part of the $\mathcal{H}_1$ becomes
\begin{widetext}
\begin{equation}
\mathcal{H}^{a}_1=I_{\tau}\otimes\Bigg(V_{A}P^{+}_{\sigma}+V_{B}P^{-}_{\sigma} \Bigg)+V_{AB}\Bigg[P^{-}_{\tau}\otimes\sigma^{\prime}_x\Bigg(\hat{z}_{\sigma},\frac{2\pi(\ell- p^{\prime})}{3}\Bigg)+P^{+}_{\tau}\otimes\sigma^{\prime}_x\Bigg(\hat{z}_{\sigma},\frac{2\pi (p-\ell)}{3}\Bigg)\Bigg].
\label{eq:Hintra.diag}
\end{equation}
\end{widetext}
The component of the potential that points along the electron's propagation direction does not backscatter since it simply shifts the longitudinal momentum and can be removed by a gauge transformation [\onlinecite{Jaffe}].  Applying the gauge transformation, we find that the component of the off-diagonal matrix elements which contributes backscattering is proportional to $V_{AB}\sin\theta_{\ell}$, where $\ell$ labels the bond where the center of the potential is positioned.  When the potential is centered in the middle of the bond $V_A=V_B$, and the diagonal matrix elements result in a scalar potential represented by an identity matrix. There is no backscattering  by a scalar potential in metallic nanotubes due to Berry's phase of the wave function under a spin rotation [\onlinecite{Ando.Berry}].  The off-diagonal intravalley matrix elements vanish when a bond-centered impurity is on a bond that is parallel to the tube circumference ($\sin\theta_{\ell}=0$).  For example, in Fig.~\ref{fig:lattice} the circumferential vector $\br{C}$ labeling an armchair $(n,n)$ tube runs parallel to the bonds labeled by a vector $\bg{\tau}_o$.  Therefore, if the center of the Gaussian is positioned in the middle of any $\bg{\tau}_o$ bond, there will be no intravalley backscattering by this local impurity for an armchair tube as labeled in Fig.~\ref{fig:lattice}.

\begin{figure}[t]
\includegraphics*[width=3.3in]{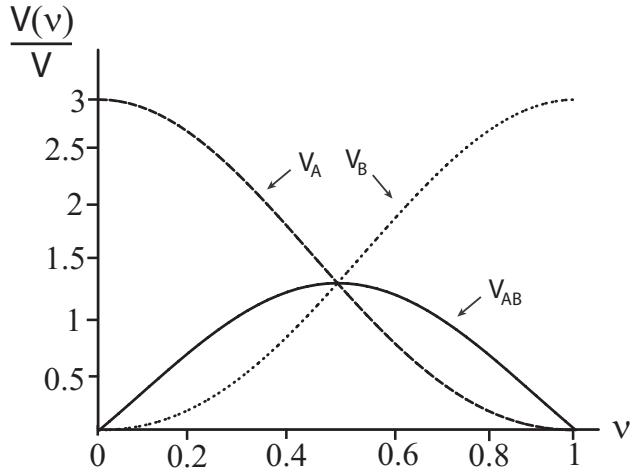}
\caption{Potential amplitudes $V_A$, $V_B$ and $V_{AB}$ defined in Eq.~(\ref{eq:coeffa}) represented by dashed, dotted and solid curves, respectively, for zero potential range $d/a\sim 0$ as a function of $\nu$.  The parameter $\nu$ determines the center of the Gaussian potential along a bond connecting two neighboring sublattice sites: $\nu=0$ indicates a potential that is $A$-sublattice centered, $\nu=1$ yields in a $B$-sublattice centered potential, and $\nu=1/2$ corresponds to a bond-centered potential.} 
\label{fig:potential.center}
\end{figure}

Bonds are parallel to the circumference only in armchair nanotubes, and a mirror reflection about the axis is accompanied by an exchange of an $A$ and $B$ sublattice.  Therefore, a mirror reflection across the nanotube axis for armchair tubes commutes with the Hamiltonian.  If a potential commutes with the Hamiltonian, left and right moving states will not mix, and there will be no backscattering.  Therefore, perturbations that are symmetric with respect to mirror reflection about the tube axis have zero intravalley backscattering amplitudes [\onlinecite{Falko,Ando2}]. 
 
In the large radius $R\gg a$ limit when the Gaussian potential becomes isotropic as shown in Eq.~(\ref{eq:2d}), the coefficients in Eq.~(\ref{eq:Hintra.diag}) within the lowest ``star'' approximation are given by
\begin{widetext}
\begin{align}
& V_A=V\Bigg\{1+\frac{2}{3}e^{-Q_o^2d^2/4}\Big[2\cos\Big(\frac{2\pi\nu}{3}\Big)+\cos\Big(\frac{4\pi\nu}{3}\Big) \Big] \Bigg\} \nonumber \\
& V_B=V\Bigg\{1+\frac{2}{3}e^{-Q_o^2d^2/4}\Big[2\cos\Big(\frac{2\pi(\nu-1)}{3}\Big)+\cos\Big(\frac{2\pi(2\nu+1)}{3}\Big)\Big] \Bigg\}  \nonumber \\
&V_{AB}=\frac{2V}{3}e^{-Q_o^2d^2/4}\Bigg\{\cos\Big(\frac{\pi(2\nu-1)}{3}\Big)+\cos\Big(\frac{2\pi(2\nu-1)}{3}\Big) \Bigg\} 
,
\label{eq:coeffa}
\end{align}
\end{widetext}
where 
\begin{equation}
Q_o=|\br{K}_p-\br{K}_{p^{\prime}}|=\frac{4\pi\sqrt{3}}{3a},~~p\neq p^{\prime} 
\label{eq:q0}
\end{equation}
is the momentum transfer between equivalent Fermi points depicted in Fig.~\ref{fig:star} in the lowest ``star'' approximation.

  For short-range potentials, the matrix structure of the scattering potential is a function of the center of the Gaussian potential $\nu$.
A plot of the amplitudes in Eq.~(\ref{eq:coeffa}) as a function of potential center $\nu$ for $d/a\sim 0$ is shown in Fig.~\ref{fig:potential.center}.  The dashed, dotted, and solid curves represent $V_A$, $V_B$, and $V_{AB}$, respectively.
When the Gaussian potential is centered on the $A$ sublattice ($\nu=0$), there is no amplitude on the $B$ sublattice ($V_B=0$) and vice versa.  The off-diagonal amplitude $V_{AB}$ is zero for both $A$ ($\nu=0$) and $B$ ($\nu=1$) sublattice centered potentials and is maximum when the potential is bond-centered ($\nu=1/2$).  When the potential is centered in the middle of the bond the three amplitude are equal $V_A=V_B=V_{AB}$. For long-ranged $d/a\gtrsim 1$ potentials, the lattice structure resolution is smeared, and $\mathcal{H}_1^a$ becomes a scalar potential which does not backscatter massless fermions.

The intervalley matrix elements that describe scattering between inequivalent 
$K$ and $K^{\prime}$ points are given by
\begin{widetext}
\begin{equation}
\mathcal{H}^{e}_1=V_{A}^{\prime}\tau_x^{\prime}(\hat{z}_\tau,\phi_A)\otimes P^{+}_{\sigma}+V_{B}^{\prime}\tau_x^{\prime}(\hat{z}_\tau,\phi_B)\otimes P^{-}_{\sigma}+\frac{V^{\prime}_{AB}}{2}\bg{\tau_{+}}\otimes\Big(\bg{\sigma_{-}^{\prime}}(\hat{z}_{\sigma},-\phi^p_{AB})
+\bg{\sigma_{+}^{\prime}}(\hat{z}_{\sigma},\phi^{p^{\prime}}_{AB})\Big),
\end{equation}
\end{widetext}
where the phases are $\phi_A=\br{K}\cdot\br{R}_A^o$, $\phi_B=\br{K}\cdot\br{R}_A^o-2\pi/3(p+p^{\prime}+\ell)$, $\phi^p_{AB}=\br{K}\cdot\br{R}_A^o-2\pi/3(p-\ell)$, and $\br{K}\cdot\br{R}_A^o=2\pi/3(n_o-m_o)$. 
 The intervalley scattering coefficients $V^{\prime}_A$, $V^{\prime}_B$, and $V^{\prime}_{AB}$ in the large radius limit and the lowest ``star'' approximation are given by
\begin{widetext} 
\begin{align}
& V_A^{\prime}=\frac{V}{3}\Bigg\{2e^{-Q_1^2d^2/4}\Big[1+2\cos\Big(\frac{2\pi\nu}{3}\Big)\Big] +e^{-Q_2^2d^2/4}\Big[1+2\cos\Big(\frac{4\pi\nu}{3}\Big)\Big]\Bigg\} \nonumber \\
& V_B^{\prime}=\frac{V}{3}\Bigg\{2e^{-Q_1^2d^2/4}\Big[1+2\cos\Big(\frac{2\pi(\nu-1)}{3}\Big)\Big] +e^{-Q_2^2d^2/4}\Big[1+2\cos\Big(\frac{2\pi(2\nu+1)}{3}\Big)\Big]\Bigg\} \nonumber \\
& V^{\prime}_{AB}=\frac{V}{3}\Bigg\{e^{-Q_1^2d^2/4}\Big[-1+2\cos\Big(\frac{\pi(2\nu-1)}{3}\Big)\Big]+e^{-Q_2^2d^2/4}\Big[1+2\cos\Big(\frac{2\pi(2\nu-1)}{3}\Big)\Big] \Bigg\}.
\label{eq:coeffe}
\end{align}
\end{widetext}
Within the lowest ``star'' there are two magnitudes of momentum transfer between distinct Fermi points which are given by 
\begin{align}
& Q_1=|2\br{K}_p|=\frac{4\pi}{3a} \nonumber \\
& Q_2=|\br{K}_p+\br{K}_{p^{\prime}}|=\frac{8\pi}{3a},~~p\neq p^{\prime}.
\label{eq:q12}
\end{align}
Intervalley amplitudes are equal to their corresponding intravalley amplitudes for atomically sharp potentials when $d/a\sim 0$.

\begin{figure}[h]
\includegraphics*[width=3.3in]{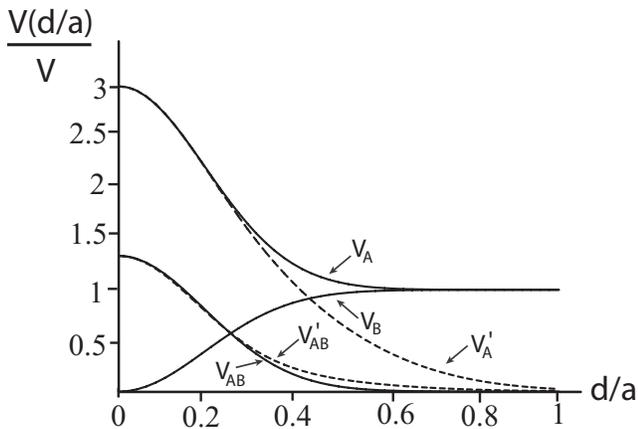}
\caption{Intravalley and intervalley amplitudes given Eq.~(\ref{eq:coeffa}) and  Eq.~(\ref{eq:coeffe}), respectively, due to a local Gaussian potential as a function of range $d/a$ for various values of potential center $\nu$.  The plots labeled $V_A$, $V_B$, and $V^{\prime}_A$ are due to a $A$-sublattice centered potential ($\nu=0$).  In this case,  $V^{\prime}_B$, $V_{AB}$, and $V^{\prime}_{AB}$ are zero.  For $d/a\gtrsim 1$, $V_A=V_B\sim V$ and $V^{\prime}_A\to 0$.  The remaining curves labeled $V_{AB}$ and $V^{\prime}_{AB}$ are off-diagonal amplitudes due to a bond-centered potential ($\nu=1/2$), which decay to zero for a long-range potential.} 
\label{fig:range}
\end{figure}

The intervalley amplitudes approach zero for long-range potentials $d/a\gtrsim 1$, unlike the diagonal intravalley terms in Eq.~(\ref{eq:coeffa}) which approach a constant.
Intravalley and intervalley amplitudes given in Eq.~(\ref{eq:coeffa}) and Eq.~(\ref{eq:coeffe}), respectively, are plotted as a function of potential range $d/a$ in Fig.~\ref{fig:range}.  The curves labeled 
$V_A$, $V_B$, and $V^{\prime}_A$ are amplitudes of a $A$-sublattice centered ($\nu=0$) potential.
Due to three-fold rotational symmetry of the lattice $V^{\prime}_B=0$ for a $A$-sublattice centered potential.  When $d/a\sim 0$ the amplitudes for intravalley and intervalley scattering become equal $V_A=V^{\prime}_A=3V$, and $V_B=0$. The vice versa is true for a $B$-sublattice centered ($\nu =1$) scatterer. Off-diagonal intravalley and intervalley amplitudes $V_{AB}$ and $V^{\prime}_{AB}$ vanish for a sublattice centered potential.  The remaining two curves are plots of off-diagonal amplitudes due to a potential centered in the middle of a bond ($\nu=1/2$).  In general, the intervalley amplitudes decays slower than the intravalley ones, since $Q_1<Q_o$.  When the potential is anisotropic for $R\sim a$, the relative magnitude of the intervalley and intravalley amplitudes is a function of the tube's chiral angle.

To summarize, for a Gaussian model potential intervalley scattering amplitudes decay as a function of $d/a$ for all values of $\nu$ and are negligible for a long-range potential.  The intravalley components of a local potential Hamiltonian also do not contribute to scattering when the potential is long-ranged.
When the range of the potential is on the order of interatomic spacing $d/a\gtrsim 1$,  the potential in Eq.~(\ref{eq:Hlocal}) becomes a scalar and is described by an identity matrix $I_{\tau}\otimes I_{\sigma}$, which produces no scattering for massless Dirac fermions [\onlinecite{Ando.Berry}]. This holds for all values of $\nu$, since the position of the potential is irrelevant when the potential is slowly varying on the scale of the lattice.  Therefore, only atomically sharp local potentials produce backscattering, a regime where both intra- and inter- valley scattering play a role.
Note, within our model one cannot realize a local potential where only intravalley scattering is present.  Therefore, a local potential inevitably results in a two-valley problem.  

\subsection{Non-Local Potentials} \label{sec:non-local}
In this section we present an example of a one-body non-local potential and show that it backscatters even when the potential is long-ranged.  We model a non-local potential by
\begin{align}
V(\br{r},\br{r^{\prime}})=V\Big(\frac{\br{r}+\br{r^{\prime}}}{2}\Big)\Big[&g(\br{r}-\br{r^{\prime}})\delta(\br{r}-\br{r^{\prime}}-\bg{\tau}_j)\nonumber \\
&+g(\br{r^{\prime}}-\br{r})\delta(\br{r}-\br{r^{\prime}}+\bg{\tau}_j) \Big].
\label{eq:nonlocal}
\end{align}
The prefactor $V(\br{\bar{r}})$ depends on the average $\bar{\br{r}}=(\br{r}+\br{r^{\prime}})/2$ of the spatial coordinates, and the remaining terms depend of the difference of $\br{r}$ and $\br{r^{\prime}}$.  The $\delta$-functions restrict the length scale of $g(\bg{\tau}_j)$ to the nearest neighbors.  The quantity $g(\bg{\tau}_j)$ can describe, for example, local modulation of the hopping integral between
neighboring sites.  This term depends on the orientation of the $j$ bond in a nanotube.   

Calculating the off-diagonal intravalley matrix element $V_{AB}$ in the lowest ``star'' we find
\begin{align}
\langle\Psi^{p}_A(\br{r})|V(\br{r})|\Psi^{p}_B(\br{r})\rangle &
=\frac{2z^p}{3}\sum_{m,m^{\prime}}V(\br{K}_{m^{\prime}}-\br{K}_m)z^{-m} \nonumber \\
&\sum_jg(\bg{\tau}_j)\cos\Big[\frac{(\br{K}_m+\br{K}_{m^{\prime}})\cdot\bg{\tau}_j}{2}\Big]. 
\label{eq:exp.nonlocal}
\end{align} 
In order to obtain the dependence of the potential on the orientation of the lattice with respect to the tube axis, we study the first three terms in the gradient expansion of $g(\bg{\tau}_j)$ given by 
\begin{equation}
\br{g}(\bg{\tau}_j)\sim g_o+\bg{\tau}_j\cdot\br{g}_1+\frac{1}{2}\bg{\tau}_j\cdot\ten{\br{g}_2}\cdot\bg{\tau}_{j},
\label{eq:expansion}
\end{equation}
where $g_o$ is a scalar, $\br{g}_{1}$ is a vector, and $\ten{\br{g}_2}$ is a tensor of rank two.  We include deviations of the hopping amplitude to zeroth order in the momentum expansion around the Brillouin zone corners.  
We fix the defect potential in the plane of a tube's coordinate system and obtain the dependence of the perturbation potential on the tube's chiral angle $\theta$.

The off-diagonal intravalley matrix elements for a non-local potential  have terms that are non-vanishing for zero momentum transfer.  We evaluate the  $m=m^{\prime}$ component of the sum in Eq.~(\ref{eq:exp.nonlocal}) for the first three terms in the gradient expansion of $\br{g}(\bg{\tau}_j)$ shown in Eq.~(\ref{eq:expansion}).   
The zeroth-order scalar $g_o$ term, the average of the hopping amplitudes, has no off-diagonal contribution at the Brillouin zone corners.  The first-order term proportional to $\br{g}_{1}$ is a vector potential that shifts the electronic spectrum around a Fermi point.  Vector potentials that couple to the longitudinal momentum have no effect on any physical properties and can be eliminated by a simple gauge transformation.  Therefore, only the the component of the vector potential that shift the momentum in the azimuthal direction can scatter incoming states.  The second-order term in the expansion couples to $\ten{\br{g}_2}$, a tensor of rank two.  These potentials describe deformations such at strains, twists, and curvature.  Some examples of such perturbations can be found in [\onlinecite{MeleKane,Kleiner}].

Including the first three terms in the gradient expansion, the dimensionless sum over $g(\bg{\tau}_j)$ that enters the $m=m^{\prime}$ term of Eq.~(\ref{eq:exp.nonlocal}) is given by
\begin{align}
&\sum_j \br{g}(\bg{\tau}_j)z^{-j}\sim \frac{a\sqrt{3}}{2}e^{-i\theta}\Big[ \gc+i\ga\nonumber \\
&+\frac{a}{4\sqrt{3}}e^{i3\theta}[(\Gcc-\Gaa)-i(\Gac+\Gca)] \Big]\equiv \tilde{g}e^{-i\theta},
\label{eq:gradient}
\end{align}
where we have used $\exp(i\br{K}_m\cdot\bg{\tau}_j)=z^{m-j}$, and $\sum_mz^{\pm m}=0$.
The components of the two-dimensional vector potential $\br{g}_1$ along the tube axis and circumference are defined by $\ga$ and $\gc$, respectively, and have dimensions of inverse length.  The vector potential does not depend on the chiral angle as seen in Eq.~(\ref{eq:gradient}).  The components of the rank two tensor $\ten{\br{g}_2}$ are defined by $G_{ij}$ with dimensions of inverse length squared.  For example, the diagonal components $\Gaa$ and $\Gcc$ can result from uniaxial strains along the axial and circumferential directions, respectively. The off-diagonal components $\Gac$ and $\Gca$ can represent strains such as local twists [\onlinecite{MeleKane},\onlinecite{Kleiner}].  The tensor potential preserves the symmetry of the honeycomb lattice since it is invariant under the transformation of the chiral angle $\theta$ by $2\pi/3$, which is apparent in the $3\theta$ dependence in Eq.~(\ref{eq:gradient}).  

Gauging away the component of the potential that couples to the longitudinal momentum, the non-local defect potential due to zero-momentum transfer is given by  
\begin{equation}
\mathcal{H}_2=V\mbox{Im}(\tilde{g})[P^{-}_{\tau}\otimes\sigma^{\prime}_y(\hat{z}_{\sigma},\theta_p^{\prime})-P^{+}_{\tau}\otimes\sigma^{\prime}_y(\hat{z}_{\sigma},-\theta_p) \Big].
\label{eq:Hnonlocal}
\end{equation}
When $V(\br{\bar{r}})$ is modeled by a Gaussian potential, all other matrix elements of a non-local potential decay $\propto\exp(-Q_i^2d^2/4)$ where $Q_i$'s are defined in Eq.~(\ref{eq:q0}) and Eq.~(\ref{eq:q12}).  Therefore, these matrix elements are parametrically smaller than the ones described in Eq.~(\ref{eq:Hnonlocal}) for non-zero $d/a$ and will not be considered further.

The perturbation Hamiltonian due to a non-local potential given in Eq.~(\ref{eq:Hnonlocal}) is independent of the impurity position $\nu$ and preserves the rotational symmetry of the lattice.  The potential is non-zero for potential ranges that exceed the scale of the lattice. The range of this potential is only limited by the envelope square barrier defined in Sec.~\ref{sec:range}.  Therefore, for a non-local potential only intravalley scattering contributes for finite range potentials, and the problem is single-valley.

\section{Force Calculation and Scattering Mechanism} \label{sec:force}
In our previous work we developed a framework for studying Casimir forces between potentials relevant for the one-valley scattering in metallic carbon nanotube [\onlinecite{first.paper}].
In this paper we discuss potentials where both intra- and inter- valley scattering are present.  In this section we review the one-valley force calculation, and then generalize the method to the two-valley scattering problem.

\subsection{One-Valley Problem} \label{sec:1valley}

In Ref.~\onlinecite{first.paper}
we employ the force operator approach to calculate Casimir forces between one-valley scattering potentials mediated by one-dimensional massless Dirac fermions.    The total Hamiltonian $\mathcal{\hat H}$ for the one-valley problem is given by
\begin{equation}
\mathcal{\hat H}=-i\hbar v_FP^{+}_{\tau}\otimes\sigma^{\prime}_x(\hat{z}_{\sigma},-\theta_p)\partial_x +V(x).
\label{eq:H1valley}
\end{equation}
The first term in Eq.~(\ref{eq:H1valley}) is the $2\times 2$ low-energy Hamiltonian expanded around the $\br{K}_p$ point, obtained by decoupling the two valleys in Eq.~(\ref{eq:Ho}). The internal structure of the scattering potential is dictated by its spinor polarization.  We study potentials with sharp walls and calculate a force as the walls becomes impenetrable.     
We model a delta-function potential by a square barrier and study limits of zero width and infinite potential strength. 
  The potential $V(x)$ is given by
\begin{equation}
V(x)=Ve^{i\sigma_x\phi /2}\sigma_ze^{-i\sigma_x\phi/2}\theta(x-x_1)\theta(x_2-x),
\end{equation}
where $\phi$ is the spinor polarization of the potential, and $\theta(x)$ is a step function.

The force operator is given by 
\begin{equation}
\hat{F}=-\frac{\partial{}\hat{\mathcal{H}}}{\partial{} \bar x},
\end{equation}
where $\bar x$ is a position.  Using the Hellmann-Feynman theorem the total force is the ground state expectation value of the force operator summed over all occupied states.  The force exerted on one barrier with sharp walls is given by
\begin{align}
F=-\int_0^{\infty}dk\Big[&\left\langle\Phi(\bar x +W/2)\right|\hat{V}\left|\Phi(\bar x +W/2)\right\rangle \nonumber \\
-&\left\langle\Phi(\bar x -W/2)\right|\hat{V}\left|\Phi(\bar x
-W/2)\right\rangle\Big], \label{eq:gen.force}
\end{align}
where $\bar{x}=(x_1+x_2)/2$ is the center of the barrier, and $W=x_2-x_1$ is its width.  The wave functions in Eq.~(\ref{eq:gen.force}) are linear combinations of right- and left- moving eigenstates of the one-valley unperturbed Hamiltonian.  The relative amplitudes of the propagating states are defined by transmission and reflection coefficients.  The scattering coefficients are obtained from the transfer matrix $\Phi(x_2)=T\Phi(x_1)$ relating the wave functions at the two boundaries of a barrier.  The two expectation values in Eq.~(\ref{eq:gen.force}) represent the difference between pressures on the right and left sides of the barrier.  For the one scatterer system the pressures on both sides of the barrier are equal, and the net force exerted on the scatterer is zero.  

\begin{figure}[h]
\includegraphics*[width=3.3in]{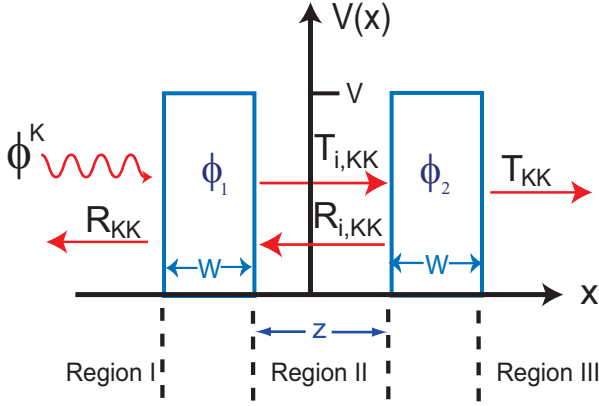}
\caption{A one-valley scattering illustration due to a $K$-point state incoming from the left.  The two barriers of width $W$ and height $V$ are separated by distance $z$.  Each barrier is characterized by its spinor polarization $\phi$.  The scattering coefficients are labeled in each region of free propagation.} 
\label{fig:potentials.1valley}
\end{figure} 

A non-zero force arises from multiple reflections of states between two or more scatterers.  A scattering process between two barriers due to a right-moving state is illustrated in Fig.~\ref{fig:potentials.1valley}. The scattering potentials are labeled by their spinor polarization $\phi$.  The reflection and transmission coefficients resulting from scattering processes within the same valley are shown in Fig.~\ref{fig:potentials.1valley}.
For example, $R_{KK}$ the amplitude of a right-moving $K$ state backscattered into a left-moving $K$ state.   

To calculate the force between two barriers, we fix the position of the left barrier and differentiate the Hamiltonian with respect to their separation $z$.  The total force is given by
\begin{equation}
F=\int_0^{\infty}\frac{dk}{2\pi}k\Bigg[2-\sum|R_{i,KK}|^2-\sum |T_{i,KK}|^2 \Bigg],
\label{eq:force.1valley}
\end{equation}
where the sum is over coefficients due to right and left incoming states.  The first term in Eq.~(\ref{eq:force.1valley}) is an outer pressure pushing the barriers together, and the remaining terms represent an inner pressure pushing the barriers apart.  In the two barrier system, the outer and inner pressures are not equal resulting in a non-zero force.  

\begin{figure}[h]
\includegraphics*[width=3.3in]{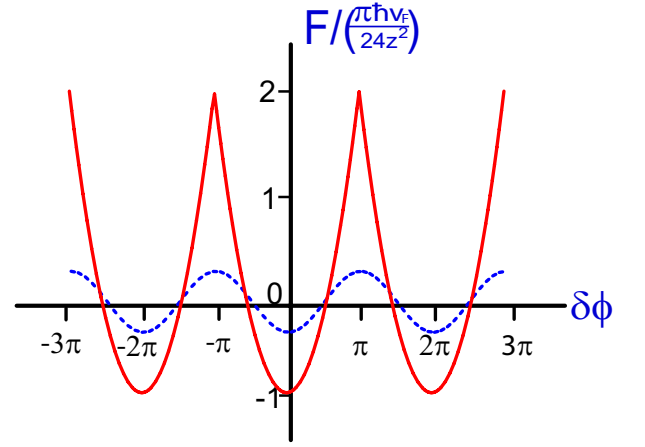}
\caption{Force between two barriers as a function of their relative spinor polarization $\delta\phi$.  The solid and dashed lines represent the forces in the large and small potential strength limits given in Eq.~(\ref{eq:F1valley}), respectively.  The magnitude of the force in the weak potential limit, the dashed curve, is rescaled to $\Gamma=1/2$ so the two curves can be compared.}
\label{fig:force.plot}
\end{figure}

We obtain a force whose sign and magnitude depends on the relative spinor polarization $\delta\phi=\phi_2-\phi_1$ of the two scatterers.  The force between two barriers separated by distance $z$ in the strong and weak strength $\Gamma=VW/\hbar v_F$ limits is given by [\onlinecite{first.paper}] 
\begin{equation}
F=-\frac{\hbar v_F\pi}{24 z^2}\left\{ \begin{array}{rl}
&1-3\Big(\delta\phi/\pi\Big)^2,~~~~~\mbox{$\Gamma\gg 1$}\\ \\
& 12\Gamma^2\cos(\delta\phi)/\pi^2,~~~\mbox{$\Gamma\ll 1$}
\end{array}
\right.
\label{eq:F1valley}
\end{equation}
In the $\Gamma\gg 1$ limit $-\pi\leq\delta\phi <\pi$ beyond which the force is periodic.
When two potentials are aligned at $\delta\phi=2\pi n$, we obtain a universal attractive force for the fermionic Casimir effect in one-dimension.  When $\delta\phi=(2n+1)\pi$ the relative spinor polarization of the two scatterers is antiparallel resulting in a repulsive force.  The oscillatory dependence on $\delta\phi$ persists in the weak strength limit. Note, it is convenient to express the force in the strong limit in terms of a dilogarithm function $Li_2(x)$, $F=\hbar v_F\mbox{Re}[Li_2(-e^{i\delta\phi})]/2\pi z^2$ in Eq.~(\ref{eq:F1valley}) when $\Gamma\gg 1$. 
 The results in Eq.~(\ref{eq:F1valley}) are plotted in Fig.~\ref{fig:force.plot}.

  In the $\Gamma\gg 1$ limit the states between the barriers are quantized, and the number of states changes by one when $\delta\phi$ is an odd multiple of $\pi$ resulting in the cusps seen in Fig.~\ref{fig:force.plot}.  The weak limit does not exhibit this behavior, since the quasibound states between the scatterers are described by a continuous spectrum.  
  
\subsection{Two-Valley Problem} \label{sec:2valley}

\begin{figure}[b]
\includegraphics*[width=3.3in]{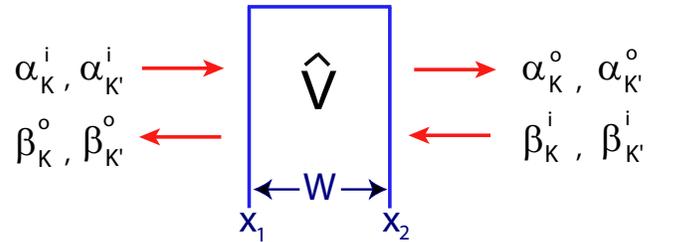}
\caption{An illustration of a scattering mechanism by a square barrier potential described by a matrix $\hat{V}$ and width $W$. A $4\times 4$ scattering matrix is obtained by relating right and left moving $K$ and $K^{\prime}$ states to their corresponding outgoing states.} 
\label{fig:scattering.matrix}
\end{figure}

In this section we generalize the method described in Sec.~\ref{sec:1valley} to the two-valley scattering problem, where scattering of states between different valleys as well as within the same valley is present.  Therefore, the potential is described by a $4\times  4$ matrix characterized by sublattice and valley degrees of freedom.  The  intra- and inter- valley matrix elements are obtained using the Bloch basis states described in Sec.~\ref{sec:basis}.  The freely propagating states are eigenstates of the effective Hamiltonian given in Eq.~(\ref{eq:Ho}). 
The wave functions used to calculate the force expectation values obtained from the Hellmann-Feynman theorem are linear combination of right and left moving states from the two $K$ and $K^{\prime}$ points.  The relative amplitudes of the propagating states are defined by scattering coefficients.  A general expression for the wave function in a region of free propagation is given by
\begin{align}
\Phi(x)=&e^{ik\ax}\Big(\alpha_{K}\phi^{K}_k+\alpha_{K^{\prime}}\phi^{K^{\prime}}_k\Big)\nonumber\\
+&e^{-ik\ax}\Big(\beta_{K}\phi^{K}_{-k}+\beta_{K^{\prime}}\phi^{K^{\prime}}_{-k}\Big),
\end{align}
where $\phi$'s are four component spinors given in Eq.~(\ref{eq:spinors}), and $\alpha$'s and $\beta$'s are scattering coefficients.  For simplicity of notation we have dropped the $p$ and $p^{\prime}$ superscripts referring to one of the three equivalent corner points.  The three $K$ points are related by reciprocal lattice vectors, and physical quantities will not depend on the particular choice of the corner point.  The dependence on $p$ and $p^{\prime}$ enters only as a phase of the scattering coefficients $\alpha$'s and $\beta$'s.    
\begin{figure*}[t]
\includegraphics*[width=5.5in]{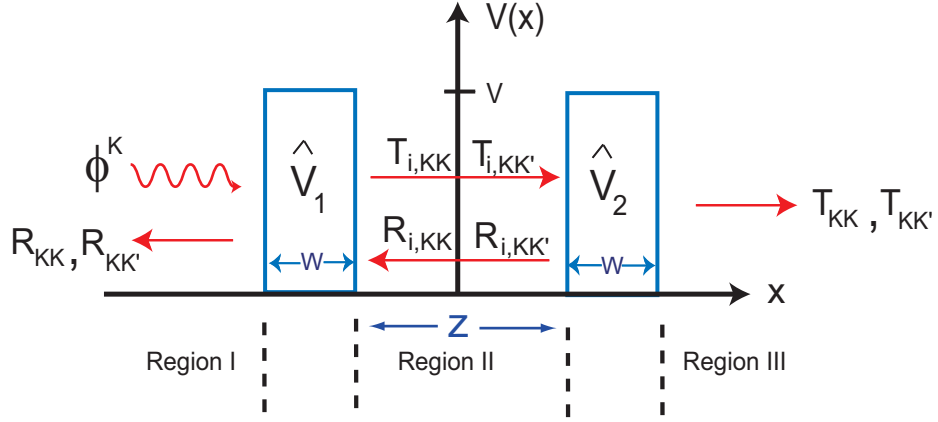}
\caption{A right-moving state $\phi^{K}_{k}$ is scattered by a two barrier system separated by distance $z$ along the tube axis.  Each barrier has a width $W$, height $V$, and is labeled by an $4\times 4$ matrix-valued potential $\hat{V}$.  Generally, each potential can produce both intravalley and intervalley scattering as labeled by the appropriate coefficients in each region of free propagation.} 
\label{fig:potentials}
\end{figure*} 

The full Hamiltonian for the one square barrier system is given by
\begin{equation}
\mathcal{\hat H}_T=\mathcal{\hat H}_o+\hat{V}\theta(\ax-x_1)\theta(x_2-\ax),
\label{eq:Ht}
\end{equation}
where $\mathcal{\hat H}_o$ is the low-energy Hamiltonian given in Eq.~(\ref{eq:Ho}), $\hat V$ is a perturbation potential, such as $\mathcal{H}_1$ or $\mathcal{H}_2$ described in Sec.~\ref{sec:H1}, and the step functions define a square barrier.
Integrating Eq.~(\ref{eq:Ht}) across the barrier, the $4\times 4$ transfer matrix becomes
\begin{align}
T=\exp\Big\{-iW[&P^{+}_{\tau}\otimes\sigma^{\prime}_x(\hat{z}_{\sigma},-\theta_p)\nonumber \\
-&P^{-}_{\tau}\otimes\sigma^{\prime}_x(\hat{z}_{\sigma},\theta_p^{\prime})][k+\hat{V}]\Big\},
\end{align}
where $W=x_2-x_1$ is the barrier width.

From the transfer matrix, we calculate the scattering matrix. The $4\times 4$ scattering matrix, obtained from incoming and outgoing states illustrated in Fig.~\ref{fig:scattering.matrix}, is defined as
\begin{equation}
\begin{pmatrix}
\alpha^{o} \\ \beta^{o} \\
\end{pmatrix}
=
\begin{pmatrix}
t & r^{\prime} \\ r & t^{\prime} \\
\end{pmatrix}
\begin{pmatrix}
\alpha^{i} \\ \beta^{i} \\
\end{pmatrix}
,
\label{eq:Smatrix}
\end{equation}
where $\alpha^{o(i)}=(\alpha_{K}^{i(o)},\alpha_{K^{\prime}}^{i(o)})^{T}$ are right-moving incoming ($i$) and outgoing ($o$) amplitude column vectors, and $\beta$'s define left-moving states as shown in Fig.~\ref{fig:scattering.matrix}. The ``primes'' in Eq.~(\ref{eq:Smatrix}) indicate the coefficients due to the states incoming from the right.  
 Each coefficient in the scattering
matrix in Eq.~(\ref{eq:Smatrix}) is a $2\times 2$ matrix defining both intravalley and intervalley scattering amplitudes.  For example,
\begin{equation}
t=
\begin{pmatrix}
t_{KK} & t_{K^{\prime}K}\\ t_{KK^{\prime}} & t_{K^{\prime}K^{\prime}} \\
\end{pmatrix}
,
\end{equation}
where the diagonal(off-diagonal) terms are the intravalley(intervalley) transmission coefficients. For instance, $t_{KK^{\prime}}$ is the forwardscattering amplitude of a right-moving $K$ state  being transmitted into a right-moving $K^{\prime}$ state.

As in the one-valley problem, non-zero forces arise from interactions between two scatterers. An scattering process illustration of a left-incoming $K$ state between two potentials $\hat{V}_1$ and $\hat{V}_2$ separated by distance $z$ along the tube axis is shown in Fig.~\ref{fig:potentials}.  As before, we fix the left barrier and calculate the force exerted on the right barrier using the Hellmann-Feynman theorem.  The force is given by
\begin{equation}
F=\int_0^{\infty}\frac{dk}{2\pi}k\Big[4-\sum |T_i|^2-\sum |R_i|^2 \Big],
\label{eq:force.gen}
\end{equation} 
where the summations represent a sum over all reflection and transmission coefficients in-between the two barriers (region II in Fig.~\ref{fig:potentials}) due to right and left incoming states, $\phi^{K}_{\pm k}$ and $\phi^{K^{\prime}}_{\pm k}$.  Throughout this paper lower-case coefficients will refer to scattering by one barrier, and upper-case ones due to scattering by a two barrier system.

The first term in Eq.~(\ref{eq:force.gen}) represents an outer pressure in Regions III of Fig.~\ref{fig:potentials} due to a continuous spectrum of states pushing the barriers together.
The second and third terms in Eq.~(\ref{eq:force.gen}) result in the inner pressure pushing the barriers apart, which is obtained from the coefficients in Region II of Fig.~\ref{fig:potentials}.  These coefficients are given by
\begin{eqnarray}
T_i=t_1+t_1^{\prime}(1-r_2r_1^{\prime})^{-1}r_2t_1 \nonumber \\
T_i^{\prime}=t_2^{\prime}+r_2(1-r_1^{\prime}r_2)^{-1}r_1^{\prime}t_2^{\prime}\nonumber\\
R_i=r_2(1-r_1^{\prime}r_2)^{-1}t_1 \nonumber\\
R_i^{\prime}=r_1^{\prime}(1-r_2r_1^{\prime})^{-1}t_2^{\prime}. \nonumber\\
\label{eq:S2}
\end{eqnarray}
When the intervalley matrix elements are zero in one of the scattering potentials $\hat V$, there is no forward- and back- scattering between inequivalent Fermi points for the two-barrier system.  In this case Eq.~(\ref{eq:force.gen}) reduces to the one-valley force given in Eq.~(\ref{eq:force.1valley}).

\section{Results} \label{sec:results}

Using the method described in Sec.~\ref{sec:force},
we explore the dependence of the force between two scatterers on the matrix structure, range, and strength of the defect potentials.  We distinguish interactions between local and non-local potentials discussed in Sec.~\ref{sec:H1}.
We show that the Casimir force decays as $1/z^2$ which is a universal result in one-dimension in the far field limit.  However, we also find that in the presence of intervalley scattering there is a spatially periodic modulation of this force.  Our results pertain to the limit $z\gg W$ where shape corrections are negligible [\onlinecite{first.paper}].  
A general solution of the integrals appearing in the force calculations in derived in the Appendix, and a summary of our results is presented in Table~\ref{tab:forces}. 

\subsection{Forces between Local Potentials} \label{sec:Flocal}
In this section we first consider interactions between local potentials.  As discussed in Sec.~\ref{sec:local}, backscattering from a local potential is significant for potential that vary on the scale of the lattice $d/a\lesssim 1$.  Let us specialize  Eq.~(\ref{eq:Hlocal}) to describe impurities that are centered at either of the two sublattice sites.  We first study the strong potential limit by fixing the area of the potential $\Gamma=VW/\hbar v_F$. The force is independent of the magnitude of the potential in the $\Gamma\gg 1$ limit and is relevant for the discussion of universal Casimir interactions.  For a sublattice centered potential in the atomically sharp limit $d/a\to 0$ intra- and inter- valley amplitudes are equal $V_i\sim V^{\prime}_i$, as shown in Fig.~\ref{fig:range}. All reflection and transmission coefficients for such scatterers approach the same value in the strong potential limit, $|r_{ij}|=|t_{ij}|=1/2$ $\forall~\{i,j\}=\{K,K^{\prime}\}$ and are independent of the sign of the potential. 

\begin{figure*}[t]
\includegraphics*[width=4in]{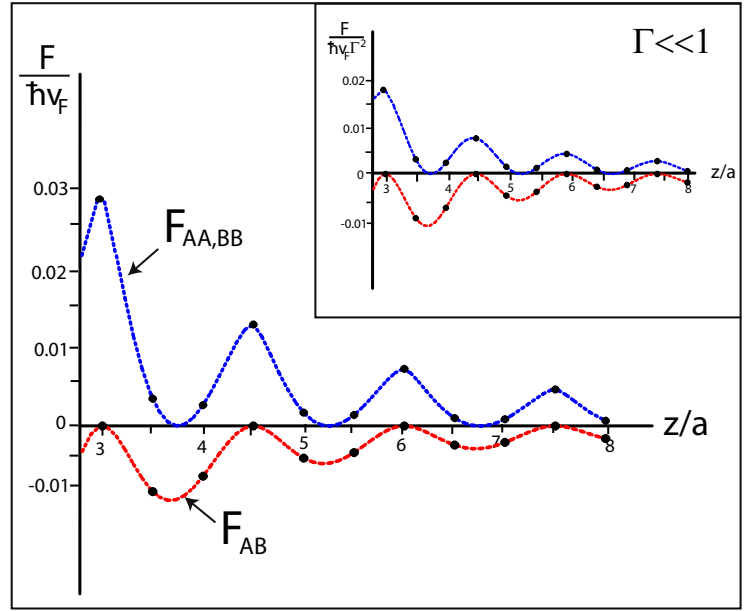}
\caption{Forces between sublattice centered impurities as a function of position.  The force $F_{AA,BB}$ between equivalent impurities given in Eq.~(\ref{eq:f.AA}) and $F_{AB}$ between defects residing on different sites given in Eq.~(\ref{eq:f.AB}) is plotted as a function of $z/a$ for an armchair tube in the strong potential limit.  The continuous limits of the force functions are shown by dashed curves in order to stress the periodicity of the spatial modulation of the forces.  The points indicate the discrete values of the force.  The inset shows equivalent results in the weak potential strength limit given in Eq.~(\ref{eq:Flocal.small}).} 
\label{fig:Fboth}
\end{figure*}

Calculating the two-barriers scattering coefficients described in Eq.~(\ref{eq:S2}) and inserting into Eq.~(\ref{eq:force.gen}), the force between two impurities centered on equivalent sublattice sites is given by
\begin{widetext}
\begin{equation}
F_{AA,BB}=\frac{\hbar v_F}{\pi}\int_{0}^{\infty}kdk\Bigg[1-\frac{1-\cos^4(\br{K}\cdot\br{R}_o)}{1+\cos^4(\br{K}\cdot\br{R}_o)-2\cos^2(\br{K}\cdot\br{R}_o)\cos(2kz)}\Bigg], 
\label{eq:force.AA}
\end{equation}
\end{widetext}
where $\br{R}_o$ is a primitive translation vector in the tangent plane separating the two impurities, and $z$ the component of their separation along the axial direction.  The subscripts $AA$ and $BB$ imply a force between impurities which are located on equivalent sites.
Applying Eq.~(A13), the solution of the force integral in Eq.~(\ref{eq:force.AA}) is given by
\begin{equation}
F_{AA,BB}=\frac{\hbar v_F}{2\pi z^2}Li_2\Big[\cos^2(\br{K}\cdot\br{R}_o)\Big]. 
\label{eq:f.AA}
\end{equation}
Unlike in the one-valley problem where the force decays monotonically as $1/z^2$, in addition the two-valley problem results in a spatial modulation of the force, as observed in the argument of the dilogarithm function in Eq.~(\ref{eq:f.AA}). 
The force oscillates with the period of the $\sqrt{3}\times\sqrt{3}$ superlattice indicating coupling between the two valley points.  The force given by Eq.~(\ref{eq:f.AA}) is plotted in Fig.~\ref{fig:Fboth} as a function of $z/a$ for an armchair tube.  The points on the curve indicate the discrete values of the force in each period.  The force between two equivalent impurities is purely repulsive, as seen in Fig.~\ref{fig:Fboth}, since $Li_2\Big[\cos^2(\br{K}\cdot\br{R}_o)\Big]>0$, where $\cos^2(\br{K}\cdot\br{R}_o)=\{1,1/4\}$.

Next, we consider interactions between impurities residing on different sublattice sites.
A force between an $A$-centered ($\nu=0$) and a $B$-centered ($\nu=1$) scatterer is given by
\begin{widetext}
\begin{equation}
F_{AB}=\frac{\hbar v_F}{\pi}\int_{0}^{\infty}kdk\Bigg[1-\frac{1-\sin^4(\br{K}\cdot\br{R}_o+\theta)}{1+\sin^4(\br{K}\cdot\br{R}_o+\theta)+2\sin^2(\br{K}\cdot\br{R}_o+\theta)\cos(2kz)}\Bigg],
\label{eq:force.AB}
\end{equation}
\end{widetext}
where $\theta$ is the chiral angle of a nanotube.
Applying Eq.~(A13) the force in Eq.~(\ref{eq:force.AB}) becomes
\begin{equation}
F_{AB}=\frac{\hbar v_F}{2\pi z^2}Li_2\Big[-\sin^2(\br{K}\cdot\br{R}_o+\theta)\Big].
\label{eq:f.AB}
\end{equation}
For unlike impurities the force is purely attractive for all values of the chiral angle.
The argument of the dilogarithm takes three values $\sin^2(\br{K}\cdot\br{R}_o+\theta)=\{\sin^2(\theta),\sin^2(2\pi/3+\theta),\sin^2(4\pi/3+\theta)\}$ which also contains $\sqrt{3}$ periodicity. The force given in Eq.~(\ref{eq:f.AB}) is plotted in Fig.~\ref{fig:Fboth} on a curve labeled $F_{AB}$ for an armchair tube as a function of position. 

\begin{figure}[h]
\includegraphics*[width=3.2in]{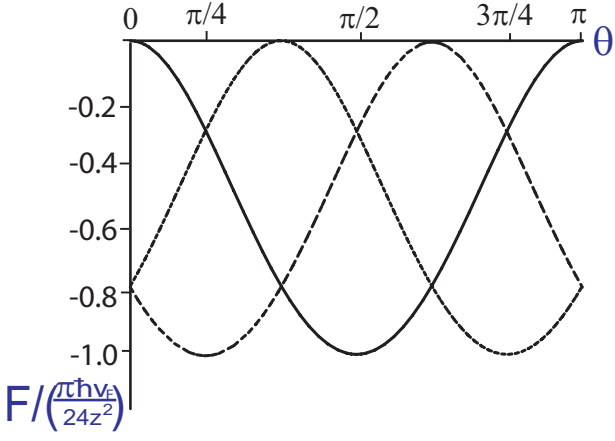}
\caption{The three branches in one period of $F_{AB}$, a force between an $A$ and a $B$ sublattice centered impurities, given in Eq.~(\ref{eq:f.AB}) as a function of chiral angle $\theta$.  The force is scaled by a factor of $\pi\hbar v_F/24z^2$ and is found to be attractive for all values of $\theta$.} 
\label{fig:chiral}
\end{figure}

Eq.~(\ref{eq:f.AB}) indicates that   
the system is invariant under the rotation of the chiral angle by $\pi$, rather than by $2\pi/3$ as for a defect-free lattice.  This occurs because the impurity is fixed on the lattice rather than on the tube's coordinates, and the position of the scatterer co-rotates with the lattice for various values of the chiral angle.  Therefore, the three-fold symmetry in the presence of an atomically sharp impurity is broken. The chiral angle dependence appears only in the force between unlike impurities, since the separation between the two defects is not a primitive lattice vector.    The three branches in one period of $F_{AB}$  are plotted 
as a function of $\theta$ in Fig.~\ref{fig:chiral}.  The figure indicates that force oscillates between $0$ and $-\pi\hbar v_F/24z^2$ for all values of $\br{K}\cdot\br{R}_o$.  An attractive and repulsive interaction between defects on different and same sublattice sites, respectively, was recently shown in two-dimensional graphene [\onlinecite{Levitov}]. 
\begin{figure*}[t]
\includegraphics*[width=4in]{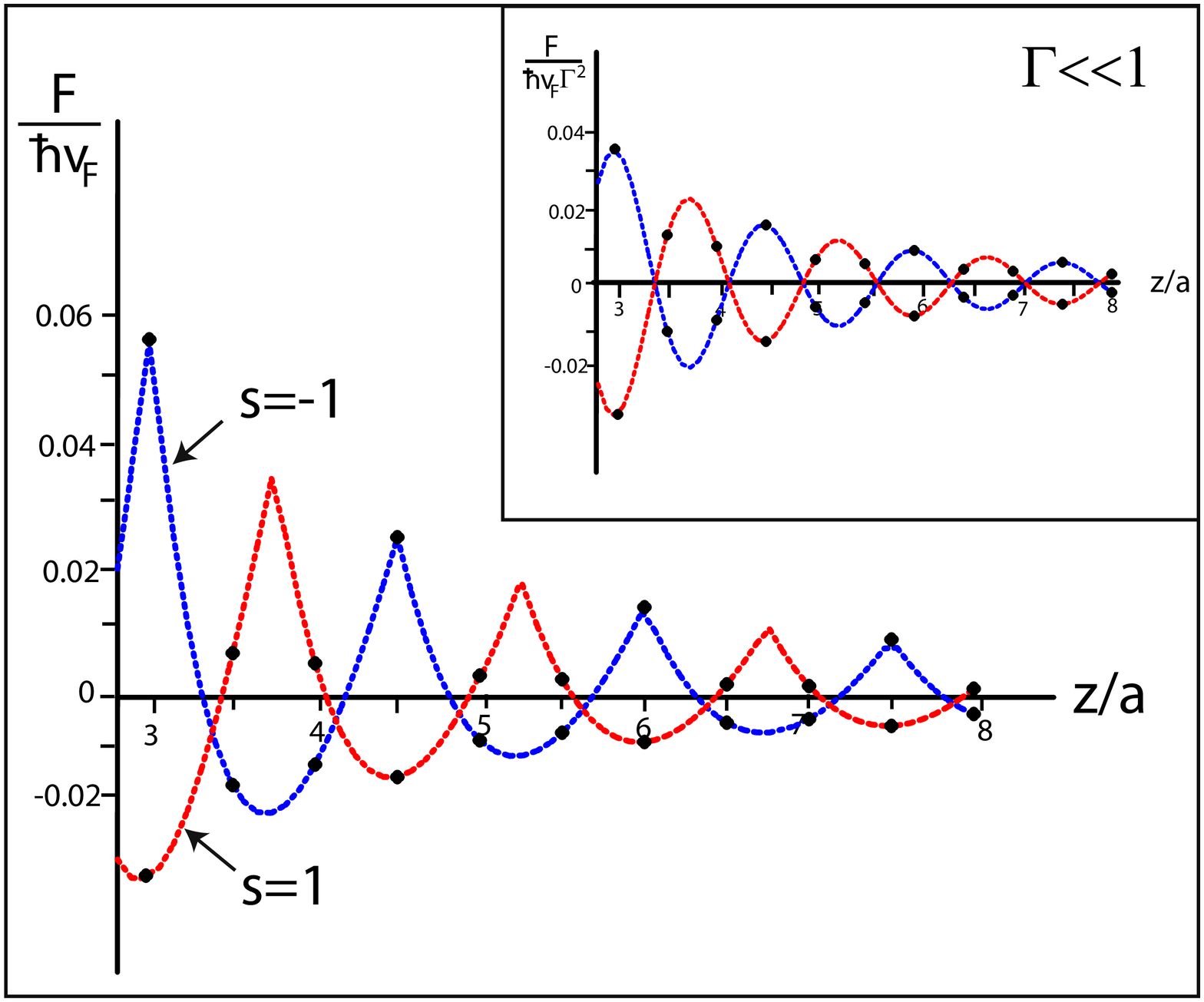}
\caption{Forces between local impurities where only intervalley scattering is present.  The force $F^e$ given in Eq.~(\ref{eq:force.inter}) between two potentials of equal ($s=1$) and unequal ($s=-1$) signs is plotted as a function of $z/a$ for an armchair tube.  The continuous limits of the force functions are shown by dashed curves.  The points indicate the discrete values of the force.  The inset shows equivalent results in the weak potential strength limit given in Eq.~(\ref{eq:Finter.small}).} 
\label{fig:Finter}
\end{figure*} 

Next, we study the small potential $\Gamma\ll 1$ limit and compare results to the ones obtained in strong $\Gamma\gg 1$ limit given by  Eq.~(\ref{eq:f.AA}) and Eq.~(\ref{eq:f.AB}).  We keep the first non-zero term in the expansion of small $\Gamma$ and take the zero width limit $W\to 0$. The next order term in the small width expansion accounting for shape corrections is  $\mathcal{O}(W/r)$ [\onlinecite{first.paper}].  For simplicity, we study the case of armchair nanotubes $\theta=0$ and find a general expression for a force between sublattice centered defects.  The off-diagonal matrix elements $V_{AB}$ and $V^{\prime}_{AB}$ are zero for sublattice centered potentials $\nu=\{0,1\}$.  The force between two local potentials 
in the $\Gamma\ll 1$ limit is given by 
\begin{align}
F=-\frac{s\hbar v_F}{4\pi z^2}\Bigg[&\Big(\Gamma^1_A-\Gamma^1_B\Big)\cdot\Big(\Gamma^2_A-\Gamma^2_B\Big)\nonumber \\
+&\Big(\Gamma^{\prime 1}_A+\Gamma^{\prime 1}_B\Big)\cdot\Big(\Gamma^{\prime 2}_A+\Gamma^{\prime 2}_B\Big)\cos(2\br{K}\cdot\br{R}_o) \Bigg],
\label{eq:Flocal.small}
\end{align}

where $s=1(-1)$ refers to a force between potentials of the same(different) sign of $\Gamma$, and the superscripts indicate the potential describing scatterer one and two. 
 Unlike in the large strength limit shown in Eqns.~(\ref{eq:force.AA})-(\ref{eq:f.AB}),  the sign of the force is a function of the relative sign $s$ of the two potentials in the weak limit.  The sign of the force also depends on the relative sublattice centers of the two scatterers, as in the strong potential limit. Therefore, in the $\Gamma\ll 1$ limit the sign of the force is controlled both by the sublattice position of the two defects and the relative sign $s$ of their potential strength.  
The $\sqrt{3}\times\sqrt{3}$ periodic oscillation persists in the small strength limit.  These results for specific sublattice positions of the two potentials and general chiral angle are shown in Table~\ref{tab:forces} and are plotted as an inset in Fig.~\ref{fig:Fboth} for an armchair tube. For long-range $d/a\gtrsim 1$ potentials the force approaches zero for all values of $\nu$ since the sublattice intravalley matrix elements $\Gamma_A$'s and $\Gamma_B$'s become equal, and intervalley terms $\Gamma^{\prime}_A$'s and $\Gamma^{\prime}_B$'s decay to zero as shown in Fig.~\ref{fig:range}.  This result confirms the absence of backscattering from an scalar potential by massless Dirac fermions.

Although a scatterer where the two valleys are decoupled cannot be realized for a local potential, a case of pure intervalley scattering $\it{is}$ possible. 
For a local potential, when an impurity is centered in the middle of a bond that points along the circumference, the potential scatters states only between inequivalent valleys as discussed in Sec.~\ref{sec:local}.  This holds because the intravalley part of the Hamiltonian $H_1^a$ is a scalar potential for all values of $d/a$, since $V_A=V_B$ for a bond-centered potential, and $V_{AB}=0$ when the perturbed bond points along the circumference.  The intervalley amplitudes are equal $V^{\prime}_A=V^{\prime}_B=V^{\prime}_{AB}$ for $\nu=1/2$, when $\theta=0$ and $\ell=0$.
 
In this case, the  intervalley transmission coefficients $|t_{KK^{\prime}}|=|t_{K^{\prime}K}|=0$ and the intravalley reflection coefficients $|r_{KK}|=|r_{K^{\prime}K^{\prime}}|=0$ vanish.  The absence of back- and forward- scattering within the same valley and between different valleys, respectively, by potentials that preserve mirror reflection symmetry about the tube axis has been also shown by Ando $\it{et~al.}$ [\onlinecite{Ando3}]. In the $\Gamma\gg 1$ limit, the non-zero coefficients have limits $|r_{KK^{\prime}}|(|r_{K^{\prime}K}|)\to 1$ and $|t_{KK}|(|t_{K^{\prime}K^{\prime}}|)\to 0$.  The phase of the reflection coefficients depends of the sign of $\Gamma$.  The force between two potentials with only intervalley scattering contribution in the large potential strength limit is given by
\begin{widetext}
\begin{equation}
F^{e}=\frac{2\hbar v_F}{\pi}\int_{0}^{\infty}kdk\Bigg[1-2\lim_{\tau\to 0}\frac{\tau^2}{|1+s(1-\tau^2)e^{2i(kz-\br{K}\cdot\br{R}_o)}|^2}\Bigg],
\label{eq:force.inter}
\end{equation}
\end{widetext}
where $\tau$ is the magnitude of the transmission coefficient.  The second term in the integrand representing the inner pressure is fundamentally different from the ones seen in Eq.~(\ref{eq:force.AA}) and Eq.~(\ref{eq:force.AB}).  The phase that appears in Eq.~(\ref{eq:force.inter}) is associated with large momentum backscattering.  The forces shown in Eq.~(\ref{eq:force.AA}) and Eq.~(\ref{eq:force.AB}) involve two types of momentum transfer which appear as various terms in the equations.
When both intra- and inter- valley play a role, there is finite transmission even in the strong potential limit.  When only intervalley scattering is present, the strong potential limit results in an impenetrable wall limit since transmission coefficient approaches zero.  Therefore, the inner pressure in Eq.~(\ref{eq:force.inter}) results from resonant states between the boundaries.  The overall prefactor in Eq.~(\ref{eq:force.inter}) is twice the magnitude than in Eq.~(\ref{eq:force.AA}) and Eq.~(\ref{eq:force.AB}). 
 
Applying Eq.~(A14) and evaluating the periodic part of the force, the solution of the integral in Eq.~(\ref{eq:force.inter}) is given by
\begin{widetext}
\begin{equation}
F^{e}=\frac{\hbar v_F}{\pi z^2}\mbox{Re}\Big[Li_2(-se^{2i\br{K}\cdot\br{R}_o})\Big] =\frac{\pi\hbar v_F}{72z^2}\left\{ \begin{array}{rl}
&\{-3,1\},~\mbox{$s=1$}\\ \\
&\{-2,6\},~~\mbox{$s=-1$}
\end{array}
\right.
\label{eq:Finter}
\end{equation} 
\end{widetext}

When only intervalley scattering amplitude is present the force oscillates between attractive and repulsive with $\sqrt{3}$ period as observed in Eq.~(\ref{eq:Finter}).  The magnitude of the force is determined by the relative sign $s$ of the two potentials.  A plot of $F^e$ as a function of $z/a$ for $s=\pm 1$ is shown in Fig.~\ref{fig:Finter}.  The points in the plot indicate the discrete values in each period of oscillation given in Eq.~(\ref{eq:Finter}). 
In the small strength limit $\Gamma\ll 1$ the force becomes
\begin{equation}
F^e=-\frac{s\hbar v_F \Gamma^2}{\pi z^2}\cos(2\br{K}\cdot\br{R}_o).
\label{eq:Finter.small}
\end{equation}
The results of Eq.~(\ref{eq:Finter.small}) are shown as an inset in Fig.~\ref{fig:Finter}.
Although the prefactors of the force are different in the two limits, the oscillation between attractive and repulsive persists in both weak and strong potential limits.  Similar behavior has been observed previously in one-dimensional Fermi liquids where only large momentum backscattering is considered [\onlinecite{Recati2},\onlinecite{liquids}].  Refer to Table~\ref{tab:forces} for a compact summary of the main results presented here.
\begin{table*}[t]
\begin{center}
\begin{tabular}{|c|c|c|c|c|c|} \hline\hline
\multirow{2}{*}{Form} & \multirow{2}{*}{Range} & \multirow{2}{*}{Site 1} & \multirow{2}{*}{Site 2} & \multirow{2}{*}{Force ($\Gamma\gg 1$) (Eq.)} & \multirow{2}{*}{Force ($\Gamma\ll 1$) (Eq.)}\\ 
& & & & &  \\ \hline
\multirow{7}{*}{Local (Eq.~\ref{eq:Hlocal})} & \multirow{5}{*}{$d/a\sim 0$} & $\nu=0$ & $\nu=0$ & \multirow{2}{*}{$\frac{\hbar v_F}{2\pi z^2}Li_2[\cos^2(\br{K}\cdot\br{R}_o)]$ (\ref{eq:f.AA})} & \multirow{2}{*}{$-\frac{s\hbar v_F \Gamma^2}{2\pi z^2}\cos^2(\br{K}\cdot\br{R}_o)$ (\ref{eq:Flocal.small})} \\
& &  $\nu=1$ & $\nu=1$ & & \\ \cline{3-6}
& &  $\nu=0$ & $\nu=1$ & \multirow{2}{*}{$\frac{\hbar v_F}{2\pi z^2}Li_2[-\sin^2(\br{K}\cdot\br{R}_o+\theta)]$ (\ref{eq:f.AB})} & \multirow{2}{*}{$\frac{s\hbar v_F \Gamma^2}{2\pi z^2}\sin^2(\br{K}\cdot\br{R}_o+\theta)$ (\ref{eq:Flocal.small})} \\
& &  $\nu=1$ & $\nu=0$ & & \\ \cline{3-6}
& & $\nu=\frac{1}{2}$ & $\nu=\frac{1}{2}$ & \multirow{2}{*}{$\frac{\hbar v_F}{\pi z^2}\mbox{Re}\Big[Li_2(-se^{2i\br{K}\cdot\br{R}_o})\Big]$ (\ref{eq:Finter})} & \multirow{2}{*}{$-\frac{s\hbar v_F \Gamma^2}{\pi z^2}\cos(2\br{K}\cdot\br{R}_o)$ (\ref{eq:Finter.small})} \\ \cline{3-4}
& & \multicolumn{2}{|c|}{($\theta=0$,~$\ell=0$)} & & \\ \cline{2-6} 
&  $d/a\gtrsim 1$ & Any & Any & 0& 0\\ \hline
\multirow{4}{*}{Non-local (Eq.~\ref{eq:Hnonlocal})} & \multirow{4}{*}{$d/a\neq 0$} & $|V|>1$ & $|V|>1$ & \multirow{2}{*}{$-\frac{\pi\hbar v_F}{12 z^2}$ (\ref{eq:Fintra})} & \multirow{2}{*}{$-\frac{\hbar v_F \Gamma^2}{\pi z^2}$ (\ref{eq:F1valley})} \\  
& & $|V|<1$ & $|V|< 1$ & & \\ \cline{3-6}
& & $|V|>1$ & $|V|<1$ & \multirow{2}{*}{$\frac{\pi\hbar v_F}{6 z^2}$ (\ref{eq:Fintra})} & \multirow{2}{*}{$\frac{\hbar v_F \Gamma^2}{\pi z^2}$ (\ref{eq:F1valley})} \\  
& & $|V|<1$ & $|V|>1$ & & \\ \hline\hline
\end{tabular}
\caption{A summary of results described in Sec.~\ref{sec:Flocal} and Sec.~\ref{sec:Fnonlocal}. The first group present results of forces between local potentials.  The remaining rows show results for forces between non-local potentials, where the dependence of the force of the relative sign $s$ of the potential strength $|V|$ is stressed.}
\label{tab:forces}
\end{center}
\end{table*}

\subsection{Forces between Non-Local Potentials} \label{sec:Fnonlocal}

In this section we calculate Casimir forces between impurities described by non-local potentials given in Eq.~(\ref{eq:Hnonlocal}).  When the range of a non-local potential is $d/a\gtrsim 0$, off-diagonal intravalley matrix elements $V_{AB}$ are dominant since all other amplitudes are parametrically smaller as noted in Sec.~\ref{sec:non-local}.  Therefore, a non-local potential can result in a one-valley scattering problem discussed in Sec.~\ref{sec:1valley}. These potentials can describe modulations to the hopping amplitudes between neighboring sites.
In the absence intervalley scattering, states are scattered only within the same  $K$ point. Therefore, the scattering coefficients $|r_{KK^{\prime}}|=|r_{K^{\prime}K}|=|t_{KK^{\prime}}|=|t_{K^{\prime}K}|=0$ are zero.  Likewise, the intervalley coefficients due to states incoming from the right vanish.   Since the two Fermi points are decoupled the perturbation matrix is described by two independent $2\times 2$ matrices in the sublattice $\sigma$-space.

The control parameter we vary to study interactions between two non-local defects is the sign of the potential $V$.  We assume that the dimensionless quantities $\mbox{Im}(\tilde{g})$'s defined in Eq.~(\ref{eq:gradient}) are equal for the two barriers.  In the strong potential $\Gamma\gg 1$ limit the magnitude of the non-zero scattering coefficients approach $|r_{KK}|(|r_{K^{\prime}K^{\prime}}|)\to 1$ and $|t_{KK}|(|t_{K^{\prime}K^{\prime}}|)\to 0$. We calculate the interaction between two barriers with the same and different signs of $\Gamma=V\mbox{Im}(\tilde{g})/\hbar v_F$.  
Applying the one-valley force result given in Eq.~(\ref{eq:force.1valley}), the force between two non-local potentials becomes
\begin{widetext}
\begin{equation}
F_2=\frac{2\hbar v_F}{\pi}\int_{0}^{\infty}kdk\Bigg[1-2\lim_{\tau\to 0}\frac{\tau^2}{|1+s(1-\tau^2)e^{2ikz}|^2}\Bigg],
\label{eq:force.intra}
\end{equation}
\end{widetext}
where $s$ is the relative sign of the two potentials.  The integrands in Eq.~(\ref{eq:force.inter}) and Eq.~(\ref{eq:force.intra}) are equivalent except for the phase $\exp(2i\br{K}\cdot\br{R}_o)$ appearing in  Eq.~(\ref{eq:force.inter}).  This phase associated with large momentum backscattering is absent in Eq.~(\ref{eq:force.intra}) since there is no intervalley scattering present by potentials given in Eq.~(\ref{eq:Hnonlocal}).

  The solution to the integral in Eq.~(\ref{eq:force.intra}) is shown in the Appendix.  Applying Eq.~(A14) the force is given by 
\begin{equation}
F_2=\frac{\hbar v_F}{\pi z^2}Li_2(-s)=\frac{\pi\hbar v_F}{12z^2}\left\{ \begin{array}{rl}
&-1,~~\mbox{$s=1$}\\ \\
&2,~~\mbox{$s=-1$}
\end{array}
\right.
\label{eq:Fintra}
\end{equation} 
The result in Eq.~(\ref{eq:Fintra}) shows that there is an attractive force between two scatterers with equal sign of $\Gamma$ $(s=1)$ and a repulsive force between defects of unequal sign of $\Gamma$ $(s=-1)$.  The relative sign of $V$ is analogous to the difference between the spinor polarizations $\delta\phi$ of the two scatterers discussed in Sec.~\ref{sec:1valley}. Potentials of equal sign ($s=1$) refer to the case of parallel scatterers $\delta\phi=0$. Two potentials of opposite sign ($s=-1$), on the other hand, refer to the case of anti-parallel scatterers $\delta\phi=\pi$. 
The results in Eq.~(\ref{eq:Fintra}) are consistent with the force in the $\Gamma\gg 1$ limit of Eq.~(\ref{eq:F1valley}) [\onlinecite{first.paper}].  Likewise, $F_2$ in the $\Gamma\ll 1$ limit agrees with Eq.~(\ref{eq:F1valley}). 
The magnitude of the force is larger than the result in Eq.~(\ref{eq:F1valley}) by a factor of two since we are including fermions from the two $K(K^{\prime})$ branches of carbon nanotubes.  These results are shown in Table~\ref{tab:forces}.

Intervalley scattering becomes important for non-local potentials when $\mbox{Im}(\tilde{g})=0$ for $d/a\lesssim 1$.  A few example of such defects are a vector potential with a zero component along the tube axis ($\ga=0$), a tensor potential for armchair tubes and zero twist ($\theta=0$ and $\Gac=\Gca=0$), or a tensor potential for zigzag tube with zero uniaxial strain ($\theta=\pi/6$ and $\Gaa=\Gcc=0$).  The effect of intervalley scattering on the Casimir force is discussed in Sec.~\ref{sec:Flocal} in the context of local potentials, and the same physics apply for the case of non-local potentials.

\section{Discussion} \label{sec:dis}

Defects or impurities on a carbon nanotube can backscatter electrons either through intravalley or intervalley scattering processes. In general both channels are present with their relative strengths determined by the range and symmetry of the scattering potentials.  The models we present here provide a framework for understanding the backscattering-induced forces on these species. The signature of intervalley scattering is a spatial modulation of the scattering-induced forces. By contrast intravalley scattering mediates a force that can be either attractive or repulsive, but has a strength that decays monotonically as a function of increasing separation. 
Interestingly, in all cases where the interaction is described by a local potential,
the scattering problem is inevitably multivalley in character, and the energy and force of the species oscillate as a function of separation.

The long-range interaction between multiple scatterers might lead to complex phase structures.  It was suggested by Shytov $\it{et~al.}$  [\onlinecite{Levitov}] that interaction between adatoms absorbed on the graphene lattice can result in defect aggregation and inhomogeneities on the lattice.

The scatterers we describe in this paper can be physically realized by various atomic and molecular species adsorbed on the tube wall.  These range from covalently bound atoms and molecules [\onlinecite{graphane,oxygen}], to more weakly bound metallic species [\onlinecite{Zettl}].  The range of the scattering potential is determined by the size of the absorbed species relative to the lattice constant.  The symmetry of the potential is determined by the spatial variation of on-site energies and by the modulation in the intersite hopping amplitudes produced by these species.  

Covalently bound species provide the most natural candidates for the strongly coupled local potential models described in
section \ref{sec:local}. Here, the on-site potential barrier at an adsorbed site can be as large as 5 eV enforcing an effectively hard wall boundary condition on the electronic wave functions. In this regime the results of section \ref{sec:Flocal} can be used to provide a bound on the electron-induced force. 
For example, the maximum attractive force between two scatterers in the impenetrable wall limit leads to an interaction
energy of $E_c=-\pi\hbar v_F/12z$. With
$\hbar v_F \sim 5.4 \, {\rm eV \cdot \AA}$ for nanotubes this gives an energy of
$2.8 \, {\rm meV}$ at a range $z=50 \, {\rm nm}$. Note that its
spatial form  follows the same scaling law as the Coulomb
interaction between uncompensated charges, but it is reduced by a
factor $\pi \hbar v_F/12 e^2 \sim .1$. Thus, for charge neutral
dipoles $p=es$ whose electrostatic interactions scale as $E_{\rm
d} \sim -p^2/z^3 = -(e^2/z) \times (s/z)^2$, they are dominated by
the Casimir interaction in the far field $z \gtrsim 5s$.
Similarly, this one-dimensional Casimir interaction completely
dominates the familiar van der Waals interactions between charge
neutral species that are mediated by the fluctuations of the
exterior three dimensional electromagnetic fields.

The weak coupling limit is relevant to the interactions of less strongly bound species, such as metal atoms or molecules bound by $\pi$ stacking interactions, e.g. benzene. Here the energy scale for the local potential is more modest, of order 1 eV which, assuming a range of order a graphite lattice constant, corresponds to a dimensionless coupling parameter $\Gamma \sim 0.5$.  In this weak coupling limit $E_l=-\hbar v_F\Gamma^2/2\pi z$ a local potential  of $V\sim 1$ eV results in $0.4$ meV at a distance of $z=50$ nm. Though weaker, this interaction still decays slowly as a function of distance  ($\propto 1/z$) and will also dominate the electrostatic interaction between charge neutral dipoles in the far field. 

In this weak coupling regime, strain induced couplings, represented  by non-local scatterers can be comparable in size.  Assuming a linear scaling of intersite hopping amplitudes with bond lengths  following $dt/d\ell \sim 4 \, {\rm eV/\AA}$ a bond length change of 0.2 \AA~and a potential range on the order of the lattice constant, this gives a dimensionless potential strength of $\Gamma \sim 0.37$ and a weak coupling interaction $E_{nl}=-\hbar v_F\Gamma^2/\pi z$, we find $0.2$ meV. These are of the same order as the forces produced by local potentials in the weak coupling limit.

For adsorbate-induced potentials, it is difficult to realize a regime where the scattering is dominated by potentials with solely a nonlocal form. Thus, one concludes that intervalley scattering and a residual spatial oscillation of the force is a generic property of inter adsorbate interactions mediated by the propagating electrons. It may be possible to quench the intervalley channel by application of a magnetic field along the tube axis which would have the effect of introducing a gap at either the K or K' point and isolating the effects of intravalley scattering. We also note that strains can be engineered into these structures by application of  mechanical stresses, and this might provide an avenue for realizing the predictions of the nonlocal model.

\section{Conclusion} \label{sec:conclusion}
In this paper we show that interactions between scatterers in metallic carbon nanotubes results in a one-dimensional Casimir problem.  We generalized our previous work which includes the one-valley problem of nanotubes, to incorporate the effects of intervalley scattering.  We show that local potentials in nanotubes produce a two-valley scattering problem. The decoupling of the two valleys is not possible for a local potential since the range must be atomically sharp in order to produce finite backscattering.  Local potentials whose spatial extent is beyond the lattice constant result in scalar potentials which do not backscatter massless Dirac fermions.  Non-local potentials, on the other hand, can result in a decoupled valley scattering problem.  Intervalley scattering amplitudes are parametrically smaller for finite range non-local potentials.  Therefore, we formulate a physically realizable potential which reduces to the one-valley scattering problem.

We study forces between two scatterers mediated by the propagating electrons of metallic carbon nanotubes.
For interactions between both local and non-local potentials we find a universal $1/z^2$ power law decay for a one-dimensional Casimir force.  However, for local potentials, where intervalley scattering plays a role, we also observe a position dependent periodic modulation of the force.  The signs and magnitudes of the forces are not universal and are controlled by the internal symmetry of the scattering potentials. 

\section*{ACKNOWLEDGMENTS} 
We would like to thank J.M. Kinder and 
Philip T. Gressman for helpful discussions.
This work was supported by the Department of Energy under Grant No. DE-FG02-ER45118.

\appendix

\section{FORCE INTEGRALS} 

In this appendix we provide a derivation for the integrals that appear in the calculations of Casimir forces for $\Gamma\gg 1$.  Although, a cutoff function is introduced in order to control divergences appearing in the integral, 
we show that the final result is cutoff independent.
The class of integrals found in this paper have a general form
\begin{equation}
F=\frac{1}{\pi}\int_0^{\infty}kdk\Bigg[\frac{1-\rho^2}{1+\rho^2\pm 2\rho\cos(2kz+\varphi)} \Bigg],
\label{eq:integral}
\end{equation}
where $z$ is the impurity separation along the tube axis.
The integrand in Eq.~(\ref{eq:integral}) can be represented in terms of a Poisson kernel
\begin{equation}
P_{s\rho}(q,\varphi)=\frac{1-\rho^2}{1+\rho^2+2\rho s\cos(q+\varphi)},  
\end{equation}
where $s=\pm 1$, and $q=2kz$.  Introducing an exponential cutoff function, the integral in Eq.~(\ref{eq:integral}) becomes
\begin{equation}
F=\lim_{\mu\to 0}\frac{1}{4\pi z^2}\int_0^{\infty}qe^{-\mu q}\Big[1-P_{s\rho}(q,\varphi) \Big]dq.
\label{eq:cutoff}
\end{equation}

  Since the Poisson kernel is $2\pi$ periodic in $q$, the integral can be expressed as an infinite sum times an integral over a region of $[0,2\pi]$.  Rewriting Eq.~(\ref{eq:cutoff}) we obtain
\begin{align}
F=\lim_{\mu\to 0}\frac{1}{4\pi z^2}\int_0^{2\pi}&\Big[1-P_{s \rho}(q,\varphi) \Big]dq\nonumber\\
&\Bigg(\sum_{n=0}^{\infty}(q+2n\pi)e^{-\mu(q+2n\pi)}\Bigg).
\label{eq:integral.cutoff}
\end{align}
Expressing the sum in terms of a geometric series and separating terms constant in $q$, the series in  Eq.~(\ref{eq:integral.cutoff}) to $\mathcal{O}(\mu)$ is given by
\begin{align}
&\sum_{n=0}^{\infty}(q+2n\pi)e^{-\mu(q+2n\pi)}=-\frac{d}{d\mu}\Bigg(\frac{e^{-\mu q}}{1-e^{-2\pi\mu}}\Bigg)\nonumber \\
&=\frac{2\pi}{(1-e^{-2\pi\mu})^2}-\frac{2\pi}{1-e^{-2\pi\mu}}+\frac{q(2\pi-q)}{4\pi}+\mathcal{O}(\mu).
\label{eq:series}
\end{align}

The first two terms on the RHS of Eq.~(\ref{eq:series}) diverge in the limit $\mu\to 0$, but vanish when integrated over $q$ since
\begin{equation}
\int_{0}^{2\pi}\Big[1-P_{s\rho}(q,\varphi) \Big]dq=0.
\label{eq:zero}
\end{equation}
To verify that the above statement is true in the case of $\rho\to 1$ we express the Poisson kernel in terms of a delta function
\begin{equation}
\lim_{\rho\to 1}P_{s\rho}(q,\varphi)=2\pi\sum_{n=0}^{\infty}\left\{ \begin{array}{rl}
\delta(q-q_n), & \mbox{$s=1$}\\
\\
\delta(q-q^{\prime}_n),  & \mbox{$s=-1$}
\end{array}
\right. 
\label{eq:lim.delta}
\end{equation}
where $q_n=\pi(2n+1)-\varphi$ and $q^{\prime}_n=2\pi n-\varphi$.  Inserting Eq.~(\ref{eq:lim.delta}) into Eq.~(\ref{eq:zero}), we find that there is either one $\delta$-function in the range of integration $[0,2\pi]$ or two $\delta$-functions at the two limits of integration, each contributing half the area.
Therefore, in both cases the integral over the series of $\delta$-functions yields a factor of $2\pi$, which is consistent with the result in Eq.~(\ref{eq:zero}).  Note, in the $\rho\to 1$ limit Eq.~(\ref{eq:cutoff}) can be solved using a generalized Abel-Plana formula which provides a finite expression for a difference between an infinite integral and an infinite sum [\onlinecite{Abel}].
 
Combining the above results and noting that the third term in Eq.~(\ref{eq:series}) is cutoff independent, Eq.~(\ref{eq:integral.cutoff}) becomes 
\begin{equation}
F=\frac{1}{16\pi z^2}\int_0^{2\pi}q(2\pi-q)\Big[1-P_{s\rho}(q,\varphi) \Big]dq.
\label{eq:integral.finite}
\end{equation}
We use the following identity to solve the integral in Eq.~(\ref{eq:integral.finite}):
\begin{equation}
\frac{1}{2\pi}\int_0^{2\pi}f(x)g(x)dx=\sum^{\infty}_{n=-\infty}\hat{f}(n)\hat{g}(-n),
\label{eq:identity}
\end{equation}
where the ``hat'' indicates the Fourier series of the original function.
The Fourier series of the Poisson kernel is given by
\begin{equation}
P_{s\rho}(q,\varphi)=\sum_{n=-\infty}^{\infty}\left\{ \begin{array}{rl}
& e^{in(q+\varphi)}\rho^{|n|},~~~~~~~~~~\mbox{$s=-1$}\\
\\
& (-1)^ne^{in(q+\varphi)}\rho^{|n|},~~\mbox{$s=1$}
\end{array}
\right.
\label{eq:Pfourier}
\end{equation}
The Fourier series of the other term in Eq.~(\ref{eq:integral.finite}) is given by
\begin{equation}
q(2\pi-q)=\frac{2\pi^2}{3}-2\sum^{\infty}_{\substack{n=-\infty\\n\neq 0}}\frac{e^{inq}}{n^2}.
\label{eq:series2}
\end{equation}
Using the results from Eqns.~(\ref{eq:identity})-(\ref{eq:series2}),  Eq.~(\ref{eq:integral.finite}) becomes
\begin{equation}
F=\frac{1}{2\pi z^2}\sum_{n=1}^{\infty}\left\{ \begin{array}{rl}
& \frac{\cos(n\varphi)\rho^{n}}{n^2},~~~~~~~~~\mbox{$s=-1$}\\ 
\\
& \frac{(-1)^n \cos(n\varphi)\rho^{n}}{n^2},~~\mbox{$s=1$}
\end{array}
\right.
\label{eq:integral.final}
\end{equation}

Eq.~(\ref{eq:integral.final}) is a general result which can be applied to all the integrals encountered in this paper.  The series above can be represented in terms of dilogarithm functions.  For example,
\begin{equation}
Li_2(-s\rho)=\sum_{n=1}^{\infty}\left\{ \begin{array}{rl}
& \frac{\rho^n}{n^2},~~~~~~~~~\mbox{$s=-1$}\\ 
\\
& \frac{(-1)^n\rho^n}{n^2}, ~~\mbox{$s=1$}
\end{array}
\right.
\label{eq:Lir}
\end{equation}
and,
\begin{equation}
\mbox{Re}[Li_2(-se^{i\varphi})]=\sum_{n=1}^{\infty}\left\{ \begin{array}{rl}
& \frac{\cos(n\varphi)}{n^2},~~~~~~~~~\mbox{$s=-1$}\\ 
\\
& \frac{(-1)^n \cos(n\varphi)}{n^2}, ~~\mbox{$s=1$}
\end{array}
\right.
\label{eq:Lif}
\end{equation}
where $Li_2(x)$ is a dilogarithm function.

In Sec.~\ref{sec:Flocal} we calculate forces between two local sublattice centered impurities. The solution of Eq.~(\ref{eq:force.AA}) for interaction between defects residing on equivalent sites is Eq.~(\ref{eq:Lir}), where $\varphi=0$, with $\rho=\cos^2(\br{K}\cdot\br{R}_o)$ and $s=-1$.  
The result for the force integral in Eq.~(\ref{eq:force.AB}), applicable to interactions between impurities centered on inequivalent sites, is Eq.~(\ref{eq:Lir}) with  $\rho=\sin^2(\br{K}\cdot\br{R}_o)$ and $s=1$.

The integral in Eq.~(\ref{eq:integral}) can also be related to integral in Eq.~(\ref{eq:force.inter}) for a force between two local potentials where only intervalley scattering plays a role, and Eq.~(\ref{eq:force.intra}) for interactions between non-local potentials.  The limit of zero transmission $\tau\to 0$ is equivalent to $\rho\to 1$ in Eq.~(\ref{eq:integral}), where $\rho=\sqrt{1-\tau^2}$.  Writing Eq.~(\ref{eq:force.inter}) and Eq.~(\ref{eq:force.intra}) in a general form in terms of $\rho$ we obtain
\begin{align}
&F=\frac{1}{\pi}\int_{0}^{\infty}k\Bigg[1-2\lim_{\rho\to 1}\frac{1-\rho^2}{|1+s\rho^2e^{i(2kz+\varphi)}|^2}\Bigg]\nonumber \\
&=\frac{1}{\pi}\int_{0}^{\infty}k\Bigg[1-\lim_{\rho\to 1}\frac{1-\rho^4}{1+\rho^4+2\rho s\cos(2kz+\varphi)}\Bigg], 
\label{eq:Fex}
\end{align}
where we have ignored the prefactors.  The right-hand side of Eq.~(\ref{eq:Fex}) is equivalent to Eq.~(\ref{eq:integral}) in the limit $\rho\to 1$.  Therefore, the solution of Eq.~(\ref{eq:force.inter}) is given by Eq.~(\ref{eq:Lif}) for $\varphi=-2\br{K}\cdot\br{R}_o$.
The solution to Eq.~(\ref{eq:force.intra}) is obtained by setting $\varphi=0$ in Eq.~(\ref{eq:Lif}).


\begin{thebibliography}{19}
\expandafter\ifx\csname natexlab\endcsname\relax\def\natexlab#1{#1}\fi
\expandafter\ifx\csname bibnamefont\endcsname\relax
  \def\bibnamefont#1{#1}\fi
\expandafter\ifx\csname bibfnamefont\endcsname\relax
  \def\bibfnamefont#1{#1}\fi
\expandafter\ifx\csname citenamefont\endcsname\relax
  \def\citenamefont#1{#1}\fi
\expandafter\ifx\csname url\endcsname\relax
  \def\url#1{\texttt{#1}}\fi
\expandafter\ifx\csname urlprefix\endcsname\relax\def\urlprefix{URL }\fi
\providecommand{\bibinfo}[2]{#2}
\providecommand{\eprint}[2][]{\url{#2}}

\bibitem[{\citenamefont{Zhabinskaya et~al.}(2008)\citenamefont{Zhabinskaya,
  Kinder, and Mele}}]{first.paper}
\bibinfo{author}{\bibfnamefont{D.}~\bibnamefont{Zhabinskaya}},
  \bibinfo{author}{\bibfnamefont{J.~M.} \bibnamefont{Kinder}},
  \bibnamefont{and} \bibinfo{author}{\bibfnamefont{E.~J.} \bibnamefont{Mele}},
  \bibinfo{journal}{Phys. Rev. A} \textbf{\bibinfo{volume}{78}},
  \bibinfo{pages}{060103(R)} (\bibinfo{year}{2008}).

\bibitem[{\citenamefont{Shytov et~al.}(2009)\citenamefont{Shytov, Abanin, and
  Levitov}}]{Levitov}
\bibinfo{author}{\bibfnamefont{A.~V.} \bibnamefont{Shytov}},
  \bibinfo{author}{\bibfnamefont{D.~A.} \bibnamefont{Abanin}},
  \bibnamefont{and} \bibinfo{author}{\bibfnamefont{L.~S.}
  \bibnamefont{Levitov}}, \bibinfo{journal}{Phys. Rev. Lett.}
  \textbf{\bibinfo{volume}{103}}, \bibinfo{pages}{016806}
  (\bibinfo{year}{2009}).

\bibitem[{\citenamefont{Fuchs et~al.}(2007)\citenamefont{Fuchs, Recati, and
  Zwerger}}]{Recati2}
\bibinfo{author}{\bibfnamefont{J.}~\bibnamefont{Fuchs}},
  \bibinfo{author}{\bibfnamefont{A.}~\bibnamefont{Recati}}, \bibnamefont{and}
  \bibinfo{author}{\bibfnamefont{W.}~\bibnamefont{Zwerger}},
  \bibinfo{journal}{Phys. Rev. A} \textbf{\bibinfo{volume}{75}},
  \bibinfo{pages}{043615} (\bibinfo{year}{2007}).

\bibitem[{\citenamefont{W{\"a}chter et~al.}(2007)\citenamefont{W{\"a}chter,
  Meden, and Sch{\"o}nhammer}}]{liquids}
\bibinfo{author}{\bibfnamefont{P.}~\bibnamefont{W{\"a}chter}},
  \bibinfo{author}{\bibfnamefont{V.}~\bibnamefont{Meden}}, \bibnamefont{and}
  \bibinfo{author}{\bibfnamefont{K.}~\bibnamefont{Sch{\"o}nhammer}},
  \bibinfo{journal}{Phys. Rev. B} \textbf{\bibinfo{volume}{76}},
  \bibinfo{pages}{45123} (\bibinfo{year}{2007}).

\bibitem[{\citenamefont{DiVincenzo and Mele}(1984)}]{Mele.mass}
\bibinfo{author}{\bibfnamefont{D.~P.} \bibnamefont{DiVincenzo}}
  \bibnamefont{and} \bibinfo{author}{\bibfnamefont{E.~J.} \bibnamefont{Mele}},
  \bibinfo{journal}{Phys. Rev. B} \textbf{\bibinfo{volume}{29}},
  \bibinfo{pages}{1685} (\bibinfo{year}{1984}).

\bibitem[{\citenamefont{Kane et~al.}(2002)\citenamefont{Kane, Mele, Johnson,
  Luzzi, Smith, Hornbaker, and Yazdani}}]{first.star}
\bibinfo{author}{\bibfnamefont{C.~L.} \bibnamefont{Kane}},
  \bibinfo{author}{\bibfnamefont{E.~J.} \bibnamefont{Mele}},
  \bibinfo{author}{\bibfnamefont{A.~T.} \bibnamefont{Johnson}},
  \bibinfo{author}{\bibfnamefont{D.~E.} \bibnamefont{Luzzi}},
  \bibinfo{author}{\bibfnamefont{B.~W.} \bibnamefont{Smith}},
  \bibinfo{author}{\bibfnamefont{D.~J.} \bibnamefont{Hornbaker}},
  \bibnamefont{and} \bibinfo{author}{\bibfnamefont{A.}~\bibnamefont{Yazdani}},
  \bibinfo{journal}{Phys. Rev. B} \textbf{\bibinfo{volume}{66}},
  \bibinfo{pages}{235423} (\bibinfo{year}{2002}).

\bibitem[{\citenamefont{Ando and Nakanishi}(1998)}]{Ando}
\bibinfo{author}{\bibfnamefont{T.}~\bibnamefont{Ando}} \bibnamefont{and}
  \bibinfo{author}{\bibfnamefont{T.}~\bibnamefont{Nakanishi}},
  \bibinfo{journal}{J. Phys. Soc. Jpn.} \textbf{\bibinfo{volume}{67}},
  \bibinfo{pages}{1704} (\bibinfo{year}{1998}).

\bibitem[{\citenamefont{Mostepanenko and Trunov}(1997)}]{casimir.book}
\bibinfo{author}{\bibfnamefont{V.~M.} \bibnamefont{Mostepanenko}}
  \bibnamefont{and} \bibinfo{author}{\bibfnamefont{N.~N.}
  \bibnamefont{Trunov}}, \emph{\bibinfo{title}{The Casimir Effect and its
  Applications}} (\bibinfo{publisher}{Clarendon Press},
  \bibinfo{address}{Oxford}, \bibinfo{year}{1997}).

\bibitem[{\citenamefont{Sundberg and Jaffe}(2004)}]{Jaffe}
\bibinfo{author}{\bibfnamefont{P.}~\bibnamefont{Sundberg}} \bibnamefont{and}
  \bibinfo{author}{\bibfnamefont{R.~L.} \bibnamefont{Jaffe}},
  \bibinfo{journal}{Ann. Phys.} \textbf{\bibinfo{volume}{309}},
  \bibinfo{pages}{442} (\bibinfo{year}{2004}).

\bibitem[{\citenamefont{Ando et~al.}(1998)\citenamefont{Ando, Nakanishi, and
  Saito}}]{Ando.Berry}
\bibinfo{author}{\bibfnamefont{T.}~\bibnamefont{Ando}},
  \bibinfo{author}{\bibfnamefont{T.}~\bibnamefont{Nakanishi}},
  \bibnamefont{and} \bibinfo{author}{\bibfnamefont{R.}~\bibnamefont{Saito}},
  \bibinfo{journal}{J. Phys. Soc. Jpn.} \textbf{\bibinfo{volume}{67}},
  \bibinfo{pages}{2857} (\bibinfo{year}{1998}).

\bibitem[{\citenamefont{McCann and Fal'ko}(2005)}]{Falko}
\bibinfo{author}{\bibfnamefont{E.}~\bibnamefont{McCann}} \bibnamefont{and}
  \bibinfo{author}{\bibfnamefont{V.~I.} \bibnamefont{Fal'ko}},
  \bibinfo{journal}{Phys. Rev. B} \textbf{\bibinfo{volume}{71}},
  \bibinfo{pages}{85415} (\bibinfo{year}{2005}).

\bibitem[{\citenamefont{Ando and Akimoto}(2004)}]{Ando2}
\bibinfo{author}{\bibfnamefont{T.}~\bibnamefont{Ando}} \bibnamefont{and}
  \bibinfo{author}{\bibfnamefont{K.}~\bibnamefont{Akimoto}},
  \bibinfo{journal}{J. Phys. Soc. Jpn.} \textbf{\bibinfo{volume}{73}},
  \bibinfo{pages}{1895} (\bibinfo{year}{2004}).

\bibitem[{\citenamefont{Kane and Mele}(1997)}]{MeleKane}
\bibinfo{author}{\bibfnamefont{C.~L.} \bibnamefont{Kane}} \bibnamefont{and}
  \bibinfo{author}{\bibfnamefont{E.~J.} \bibnamefont{Mele}},
  \bibinfo{journal}{Phys. Rev. Lett.} \textbf{\bibinfo{volume}{78}},
  \bibinfo{pages}{1932} (\bibinfo{year}{1997}).

\bibitem[{\citenamefont{Kleiner and Eggert}(2001)}]{Kleiner}
\bibinfo{author}{\bibfnamefont{A.}~\bibnamefont{Kleiner}} \bibnamefont{and}
  \bibinfo{author}{\bibfnamefont{S.}~\bibnamefont{Eggert}},
  \bibinfo{journal}{Phys. Rev. B} \textbf{\bibinfo{volume}{63}},
  \bibinfo{pages}{073408} (\bibinfo{year}{2001}).

\bibitem[{\citenamefont{Ando et~al.}(1999)\citenamefont{Ando, Nakanishi, and
  Saito}}]{Ando3}
\bibinfo{author}{\bibfnamefont{T.}~\bibnamefont{Ando}},
  \bibinfo{author}{\bibfnamefont{T.}~\bibnamefont{Nakanishi}},
  \bibnamefont{and} \bibinfo{author}{\bibfnamefont{R.}~\bibnamefont{Saito}},
  \bibinfo{journal}{Microelectronic Engineering} \textbf{\bibinfo{volume}{47}},
  \bibinfo{pages}{421} (\bibinfo{year}{1999}).

\bibitem[{\citenamefont{Elias et~al.}(2009)\citenamefont{Elias, Nair,
  Mohiuddin, Morozov, Blake, Halsall, Ferrari, Boukhvalov, Katsnelson, Geim
  et~al.}}]{graphane}
\bibinfo{author}{\bibfnamefont{D.~C.} \bibnamefont{Elias}},
  \bibinfo{author}{\bibfnamefont{R.~R.} \bibnamefont{Nair}},
  \bibinfo{author}{\bibfnamefont{T.~M.~G.} \bibnamefont{Mohiuddin}},
  \bibinfo{author}{\bibfnamefont{S.~V.} \bibnamefont{Morozov}},
  \bibinfo{author}{\bibfnamefont{P.}~\bibnamefont{Blake}},
  \bibinfo{author}{\bibfnamefont{M.~P.} \bibnamefont{Halsall}},
  \bibinfo{author}{\bibfnamefont{A.~C.} \bibnamefont{Ferrari}},
  \bibinfo{author}{\bibfnamefont{D.~W.} \bibnamefont{Boukhvalov}},
  \bibinfo{author}{\bibfnamefont{M.~I.} \bibnamefont{Katsnelson}},
  \bibinfo{author}{\bibfnamefont{A.~K.} \bibnamefont{Geim}},
  \bibnamefont{et~al.}, \bibinfo{journal}{Science}
  \textbf{\bibinfo{volume}{323}}, \bibinfo{pages}{610} (\bibinfo{year}{2009}).

\bibitem[{\citenamefont{Collins et~al.}(2000)\citenamefont{Collins, Bradley,
  Ishigami, and Zettl}}]{oxygen}
\bibinfo{author}{\bibfnamefont{P.}~\bibnamefont{Collins}},
  \bibinfo{author}{\bibfnamefont{K.}~\bibnamefont{Bradley}},
  \bibinfo{author}{\bibfnamefont{M.}~\bibnamefont{Ishigami}}, \bibnamefont{and}
  \bibinfo{author}{\bibfnamefont{A.}~\bibnamefont{Zettl}},
  \bibinfo{journal}{Science} \textbf{\bibinfo{volume}{287}},
  \bibinfo{pages}{1801} (\bibinfo{year}{2000}).

\bibitem[{\citenamefont{Regan et~al.}(2004)\citenamefont{Regan, Aloni, Ritchie,
  Dahmen, and Zettl}}]{Zettl}
\bibinfo{author}{\bibfnamefont{B.~C.} \bibnamefont{Regan}},
  \bibinfo{author}{\bibfnamefont{S.}~\bibnamefont{Aloni}},
  \bibinfo{author}{\bibfnamefont{R.~O.} \bibnamefont{Ritchie}},
  \bibinfo{author}{\bibfnamefont{U.}~\bibnamefont{Dahmen}}, \bibnamefont{and}
  \bibinfo{author}{\bibfnamefont{A.}~\bibnamefont{Zettl}},
  \bibinfo{journal}{Nature} \textbf{\bibinfo{volume}{428}},
  \bibinfo{pages}{924} (\bibinfo{year}{2004}).

\bibitem[{\citenamefont{Inui}(2003)}]{Abel}
\bibinfo{author}{\bibfnamefont{N.}~\bibnamefont{Inui}}, \bibinfo{journal}{J.
  Phys. Soc. Jpn.} \textbf{\bibinfo{volume}{72}}, \bibinfo{pages}{1035}
  (\bibinfo{year}{2003}).

\end{thebibliography}
\end{document}